\documentclass[twocolumn]{aastex631}
\usepackage{times,amsmath}
\usepackage[T1]{fontenc}

\hypersetup{pdfauthor={Kamlesh Rajpurohit},
            pdftitle={Revisiting the Group-Dominant Elliptical NGC~5044 in the Radio Band: Continuum Emission and Detection of H\textsc{i} Absorption},
            pdfkeywords={Galaxy groups; Early-type galaxies; Cooling flows; Neutral hydrogen clouds},
            bookmarksnumbered=true}
\pdfoutput=1

\usepackage[utf8]{inputenc}
\usepackage{graphicx}
\usepackage{amssymb}
\usepackage{amsmath}
\usepackage{float}
\usepackage{multirow}
\usepackage{textcomp}
\usepackage{gensymb}
\usepackage{enumitem}  
\usepackage{hyperref}
\usepackage{natbib}
\usepackage{comment}
\usepackage{systeme}
\usepackage[Symbol]{upgreek}
\usepackage{xcolor}
\usepackage{xspace}
\definecolor{xlinkcolor}{cmyk}{1,1,0,0}

\newcommand{\arcm}{\hbox{$^\prime$}}

\newcommand{\chandra}{\emph{Chandra}}

\newcommand{\arcs}{\mbox{\arcm\arcm}}

\newcommand{\Msol}{\ensuremath{\mathrm{~M_{\odot}}}}

\newcommand{\s}{\ensuremath{\mbox{~s}}}
\newcommand{\ps}{\ensuremath{\s^{-1}}}

\newcommand{\km}{\ensuremath{\mbox{~km}}}

\newcommand{\kmps}{\ensuremath{\km \ps}}
\newcommand{\mJy}{\ensuremath{\mbox{~mJy}}}
\newcommand{\beam}{\ensuremath{\mbox{~b}}}
\newcommand{\pb}{\ensuremath{\beam^{-1}}}
\newcommand{\mJypb}{\ensuremath{\mJy \pb}}

\newcommand{\gtsim}{\,\rlap{\raise 0.5ex\hbox{$>$}}{\lower 1.0ex\hbox{$\sim$}}\,} 
\newcommand{\Msmbh}{\ensuremath{M_{\mathrm{SMBH}}}}
\newcommand{\Hi}{\ensuremath{\mathrm{H\textsc{i}}}}
\newcommand{\Ha}{\ensuremath{\mathrm{H\alpha}}}

\newcommand{\hi}{\textsc{H$\,$i}\xspace}

\shorttitle{Continuum emission and \hi absorption in the early type galaxy NGC~5044}
\shortauthors{Rajpurohit et al.}

\begin{document}

\title{Revisiting the Group-Dominant Elliptical NGC~5044 in the Radio Band: Continuum Emission and Detection of H\textsc{i} Absorption}

\correspondingauthor{Kamlesh Laxmi Rajpurohit}
\email{kamlesh.rajpurohit@cfa.harvard.edu}

\author[0000-0001-7509-2972]{Kamlesh Rajpurohit}
\affiliation{Center for Astrophysics $|$ Harvard \& Smithsonian, 60 Garden Street, Cambridge, MA 02138, USA}

\author[0000-0003-1078-2539]{Tirna Deb}
\affiliation{Center for Astrophysics $|$ Harvard \& Smithsonian, 60 Garden Street, Cambridge, MA 02138, USA}

\author[0000-0002-3104-6154]{Konstantinos Kolokythas}
\affiliation{Centre for Radio Astronomy Techniques and Technologies, Department of Physics and Electronics, Rhodes University, P.O. Box 94, Makhanda 6140, South Africa}
\affiliation{South African Radio Astronomy Observatory, 2 Fir Street, Observatory 7925, South Africa}

\author[0000-0002-4760-080X]{Kshitij Thorat}
\affiliation{Department of Physics, University of Pretoria, Hatfield, Pretoria 0028, South Africa}

\author[0000-0002-5671-6900]{Ewan O'Sullivan}
\affiliation{Center for Astrophysics $|$ Harvard \& Smithsonian, 60 Garden Street, Cambridge, MA 02138, USA}

\author[0000-0002-4962-0740]{Gerrit Schellenberger}
\affiliation{Center for Astrophysics $|$ Harvard \& Smithsonian, 60 Garden Street, Cambridge, MA 02138, USA}

\author{Laurence P. David }
\affil{Center for Astrophysics $|$ Harvard \& Smithsonian, 60 Garden Street, Cambridge, MA 02138, USA}

\author[0009-0007-0318-2814]{Jan M. Vrtilek}
\affil{Center for Astrophysics $|$ Harvard \& Smithsonian, 60 Garden Street, Cambridge, MA 02138, USA}

\author[0000-0002-1634-9886]{Simona Giacintucci}
\affil{Naval Research Laboratory, 4555 Overlook Avenue SW, Code 7213, Washington, DC 20375, US}

\author[0000-0002-9478-1682]{William Forman}
\affil{Center for Astrophysics $|$ Harvard \& Smithsonian, 60 Garden Street, Cambridge, MA 02138, USA}

\author[0000-0003-2206-4243]{Christine Jones}
\affil{Center for Astrophysics $|$ Harvard \& Smithsonian, 60 Garden Street, Cambridge, MA 02138, USA}

\author[0000-0003-0231-3249]{Mpati Ramatsoku}
\affil{INAF-Cagliari Astronomical Observatory, Via della Scienza 5, I-09047 Selargius (CA), Italy}
\affiliation{Centre for Radio Astronomy Techniques and Technologies, Department of Physics and Electronics, Rhodes University, P.O. Box 94, Makhanda 6140, South Africa}
\affiliation{South African Radio Astronomy Observatory, 2 Fir Street, Observatory 7925, South Africa}

\begin{abstract}
We present new MeerKAT L-band (continuum and \hi) and upgraded Giant Metrewave Radio Telescope (300-850\,MHz) observations of the archetypal cool-core group-dominant early-type galaxy NGC~5044. Our new continuum images reveal diffuse, steep spectrum ($\alpha_{0.99\,\rm GHz}^{1.56\,\rm GHz}=-1.53\pm0.6$) radio emission extending about 25~kpc around the unresolved radio core. The observed radio emission overlaps the known X-ray cavities, but is not confined to them. We also find the first direct evidence of neutral atomic gas in NGC~5044, in the form of a 3.8$\sigma$ significant two-component \hi absorption line seen against the emission of the active nucleus. The peak velocities are well correlated with the previously reported CO(2-1) absorption, but the \hi lines are moderately broader, spanning velocities from $265\,\rm \, km\,s^{-1}$  to $305\,\rm \, km\,s^{-1}$. We do not detect \hi emission, but place an upper limit of $M_{\hi}< 5.4 \times 10^{7} \, M_{\odot}$ in the central 15\arcs\ (2.2~kpc) of the galaxy. This is significantly less than the estimated molecular gas content, and implies a molecular-to-atomic mass ratio of $\gtsim$1.7:1, consistent with these gas phases forming through cooling from the hot intra-group medium. We also constrain the spin temperature to $T_{\rm spin}\leq 950\,\rm K$, indicating that the detected \hi is in the cold neutral phase. 
\end{abstract}


\section{Introduction}
\label{sec:intro}
It is now well established that heating by active galactic nuclei (AGNs) plays an important role in balancing radiative cooling in the hot atmospheres of elliptical galaxies, galaxy groups, and clusters. Cooling appears to be regulated by a feedback loop, in which cooled gas fuels the AGN, triggering the launch of relativistic jets which heat and mix the surrounding medium via a variety of mechanisms (e.g., shocks, sound waves, turbulence, uplift), thereby suppressing the cooling (see, e.g., \citealt{Churazov2001,Churazov2002} and reviews by \citealt{Fabian12,McNamaraNulsen12,Gitti12,Gaspari2020,DonahueVoit2022}). Our understanding of AGN feedback is anchored by multi-frequency observations: in the radio continuum to trace the relativistic jets and the lobes they inflate \citep[e.g.,][]{Birzanetal08,Giacintucci2011,Birzanetal20,Pasini2021,Pasini2022}; in the X-ray to measure the thermodynamic state of the hot medium and see the impact of the jets via cavities, shock fronts, and uplifted material \citep[e.g.,][]{Churazov2000,Dongetal10,Kirkpatricketal11,Randalletal15,Russelletal19,Ubertosietal23,Plsek2024}; and via emission and absorption line observations (most commonly CO and H$\alpha$) which trace the gas which cools out of the hot phase and forms the reservoir of fuel on which the AGN draws \citep[e.g.,][]{Edge2001,SalomeCombes2003,Hogan14,Pulido2018,Lakhchauraetal18,Olivares2019,Olivares2022}. 


\begin{figure*}
    \centering
    \includegraphics[width=0.7\textwidth]{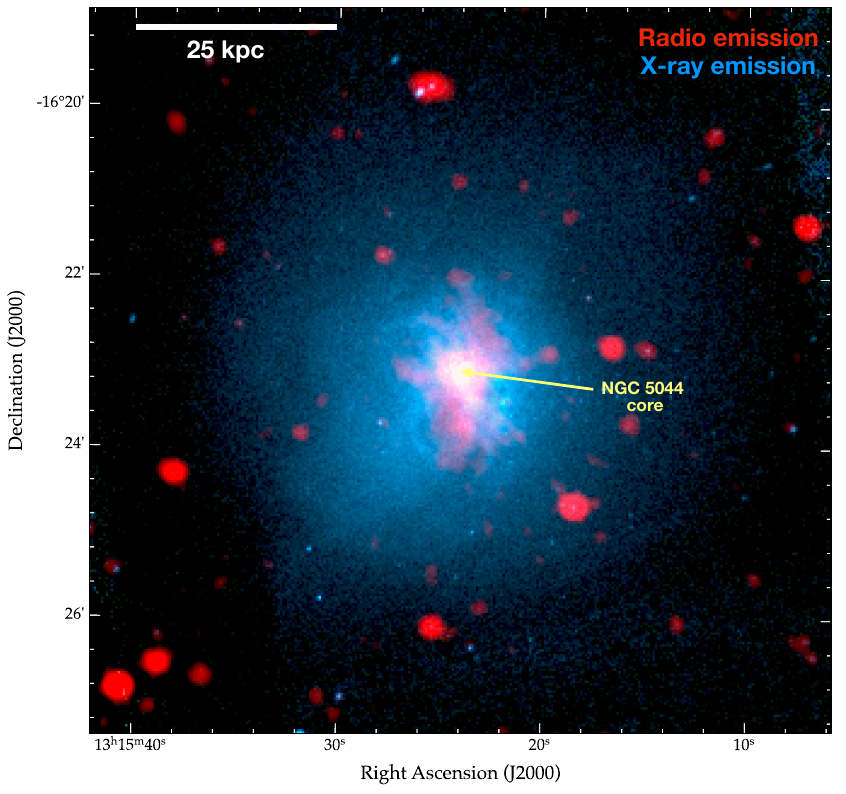}
      \vspace{-0.4cm}  
 \caption{Radio and X-ray overlay of  NGC\,5044. The intensity in red shows the radio emission observed with uGMRT at a central frequency of 380 MHz (10\arcsec\ resolution).  The intensity in blue shows \chandra\ X-ray emission in the $0.5-2.0$\,keV band.}
      \label{fig1:rgb}
\end{figure*} 

Galaxy groups are perhaps the most interesting mass scale ($\sim$10$^{13}$-10$^{14}$\Msol) for studies of the cooling/AGN feedback cycle. At the typical $\sim$1~keV temperatures of the intra-group medium (IGrM), line emission makes radiative cooling 
more efficient than in clusters, leading to the finding that there are essentially no non-cool-core groups by the standards applied to clusters \citep{OSullivanetal17}. It has also been shown that, while in clusters the gravitational binding energy of the ICM greatly exceeds the total energy available from the supermassive black holes (SMBHs) of cluster-dominant galaxies, in groups the two energies are comparable \citep{Eckert2021}. AGN feedback therefore has a much greater potential to affect the IGrM, possibly even to the point of driving gas out of the group. This is the most likely explanation for the reduced gas fractions observed in galaxy groups \citep{Gastaldello2007,Lovisari2015,Eckert2016}. Over the history of these systems, AGN feedback has heated the IGrM and driven gas out to large radii. Groups are thus a critically important environment for feedback studies, with the majority of systems likely affected by their AGN, and the feedback loop requires a fine balance to avoid expelling the IGrM.

The NGC~5044 group is the archetypal AGN feedback system for the group mass scale, with an exceptional multiwavelength observational dataset providing insight into its history and current state. X-ray observations reveal the nested cold fronts characteristic of sloshing, likely caused by the recent passage of a massive spiral galaxy through the system \citep{Buote2003a,Gastaldello2009,David2009,Gastaldello2013,O'Sullivan2014}. Nonetheless, the IGrM is highly centrally concentrated, with a steeply declining central temperature profile, short central cooling times, and low central entropy, indicating rapid cooling \citep{David1994,David2011}. The IGrM core contains a complex collection of cavities and cooling filaments \citep[see Figure~\ref{fig1:rgb};][]{David2017}. The warm and cold phases of the cooling flow are observed in the form of an extended filamentary H$\alpha$ nebula \citep{David2014} and molecular gas clouds traced by CO, [\ion{C}{2}], and rovibrational H$_2$ line emission, with CO, HCN and HCO+ also seen in absorption against the AGN \citep{Kaneda2008,David2014,Werner2014,Temi2018,Schellenberger2020,Rose2019,Rose2023,Rose2024}. The total molecular gas mass in the central 15\arcs\ (2.3~kpc) of the galaxy is estimated to be $\sim$10$^8$\Msol, one of the largest molecular reservoirs of any X-ray luminous group \citep{Schellenberger2020}. 

A combination of X-ray and radio data provides evidence of at least three cycles of feedback from the AGN of NGC~5044 \citep{Schellenberger2021}. Deep \chandra\ imaging shows a pair of cavities on $\sim$5~kpc scales, as well as a second, younger pair on sub-kiloparsec scales in the galaxy nucleus \citep{David2017}. Very Long Baseline Array observations in C and X-band (5 and 8.5~GHz respectively) show even younger parsec-scale jets \citep{Schellenberger2021}. Interestingly, these jets are oriented roughly northeast-southwest, orthogonal to the orientation of the cavities.

\setlength{\tabcolsep}{20pt}
\begin{table*}
\caption{Observational overview: MeerKAT and uGMRT observations.}
\begin{center}
\begin{tabular}{ l  c  c c c c}
  \hline  \hline  
\multirow{1}{*}{}& \multirow{1}{*}{MeerKAT} & \multicolumn{2}{c}{ uGMRT}   \\  
 \cline{3-4}
&\multirow{1}{*}{L-band} & \multirow{1}{*}{Band\,4} & \multirow{1}{*}{Band\,3}\\
\hline
Observing date & June 8, 2021 & December 2, 2023 &January 23, 2018 \\
 & -& December4, 2023&January 24, 2018 \\
Frequency coverage &0.85-1.67\,GHz& 550-950\,MHz & 300-500\,MHz\\
Channel width  &26.1\,kHz & 97\,kHz&130.2\,kHz\\ 
Number of channels  &32768 & 4096 &2048\\ 
On source time &4\,hrs &10\,hrs &4 hrs\\
\hline 
\end{tabular}
\end{center}
\label{Tabel:obs}
\end{table*} 

\setlength{\tabcolsep}{12.0pt}   
 \begin{table*}[!htbp]
\caption{Continuum image properties}
\begin{center}
\begin{tabular}{c c c c c c c r}
\hline\hline
   & Name & Restoring Beam & Robust  & \textit{uv}-cut & \textit{uv}-taper & RMS noise\\ 
&&&parameter&&&$\upmu\,\rm Jy\,beam^{-1}$\\
\hline
  \hline 
uGMRT Band3&IM1&$10\arcsec \times 10\arcsec$&$0$&$ -$&$-$&21\\
(120-168\,MHz)&IM2&$15\arcsec \times 15\arcsec$&$0$&$-$&8\arcsec&23\\
&IM3&$15\arcsec \times 15\arcsec$&$0$&$  \geq\rm0.2\,k\uplambda$&8\arcsec&25\\
\hline   
 uGMRT Band\,4 &IM4&$15\arcsec \times 15\arcsec$&$0$&$-$&$8\arcsec$&19\\
 (550-750\,MHz)&IM5&$15\arcsec \times 15\arcsec$&$0$&$ \geq\rm0.2\,k\uplambda$&8\arcsec&21\\
\hline   
MeerKAT L-band&IM6&$8.3\arcsec \times 7.6\arcsec$&$-0.5$&$-$&$-$&7\\
(0.9-1.6 GHz)&IM7&$15\arcsec \times 15\arcsec$&$0$&$-$&8\arcsec&8 \\
&IM8&$15\arcsec \times 15\arcsec$&$0$&$ \geq\rm0.2\,k\uplambda$&8\arcsec&10 \\
 
 \hline   
 \end{tabular}
 \end{center}
{Notes. Imaging was always performed in {\tt WSCLEAN} using {\tt multiscale} and with {\tt Briggs}  weighting scheme. }
\label{con_imaging}
\end{table*}

On larger scales, radio observations have presented conflicting views of the AGN activity. \citet{Romana2022} analyzed Very Large Array (VLA) L-band (1.4~GHz) radio observations, with A configuration data showed hints of kiloparsec-scale jets aligned northeast-southwest, while lower resolution AnB configuration data showing possible $\sim$4~kpc scale jets on an east-west axis. Giant Metrewave Radio Telescope (GMRT) 610~MHz data showed a possible small radio lobe extending $\sim$12~kpc southeast of the core, possibly correlated with an X-ray filament but not with the X-ray cavities \citep{Giacintucci2011}. At the lowest frequency and largest scale, a GMRT 235~MHz observation showed what appeared to be a detached radio lobe $\sim$40~kpc southeast of the core, as well as a bent radio jet connected to the radio core of NGC\,5044 \citep{Giacintucci2011}. Although no X-ray cavity was directly detected at the position of this lobe, it was found to be correlated with a region of low IGrM abundances \citep{David2011,O'Sullivan2014}, while the bends in the radio jet and the inner edge of the lobe correlated with a sloshing front, suggesting that the radio structure had been shaped by IGrM motions \citep{O'Sullivan2014}.  Of all the radio continuum observations to date, only the VLBA data show a consistent morphology at multiple frequencies, and only the GMRT 235~MHz data show a physically consistent correlation with the X-ray data.   

To try to address these inconsistencies, we have acquired deep multiband radio observations of NGC~5044 using MeerKAT at L-band (0.85-1.67~GHz) and the upgraded GMRT in Bands\,4 (550-850~MHz) and 3 (300-500~MHz). As well as offering significant increases in radio continuum sensitivity, the MeerKAT data provide an opportunity to search for neutral atomic gas in the galaxy via the \hi\ line. Given the molecular and ionized gas observed in the cool cores of groups and clusters, neutral hydrogen must be present and may be a significant mass component of the gas produced through IGrM cooling. \Hi\ has been detected in absorption in $\sim$20 brightest cluster galaxies \citep[BCGs; see, e.g.,][and references therein]{Hogan14,MorgantiOosterloo18,Sarafetal23,Rose2024} but detections in the dominant galaxies of X-ray luminous groups are rare, with only a handful of systems seen in absorption or emission  \citep[e.g.,][]{JaffeMcNamara94,Morgantietal06,Morgantietal09,Rose2024}. 

Throughout this paper, we adopt a flat $\Lambda$CDM cosmology with $H_{\rm{ 0}}=70$ km s$^{-1}$\,Mpc$^{-1}$, $\Omega_{\rm{ m}}=0.3$, and $\Omega_{\Lambda}=0.71$. We adopt a redshift-independent distance of 31.2~Mpc for NGC~5044 \citep{Tonry2001} which corresponds to a physical scale of 1\arcsec=150~pc.

All output radio images are in the J2000 coordinate system and are corrected for primary beam attenuation. We define the radio spectral index, $\alpha$, so that $S_{\nu}\propto\nu^{\alpha}$, where $S$ is the flux density at frequency $\nu$. Velocities quoted in this work are referenced with respect to the systemic heliocentric velocity of NGC\,5044 using the optical definition. Following \cite{Schellenberger2020},  we adopt a heliocentric systemic velocity of $cz = 2757\rm \,km\,s^{-1}$. All \hi uncertainties are given at the $1\sigma$ level throughout the paper.


\section{Observations and data reduction}
\subsection{MeerKAT}
We observed the NGC\,5044 with MeerKAT in the L-band (Project code: SCI-20210212).  An overview of the observation is given in Table\,\ref{Tabel:obs}. All four polarization products were recorded using the 32K correlator mode, covering a total bandwidth of 856 MHz with 32768 channels that are each 26.1 kHz wide. This corresponds to a velocity resolution of 5.5\,km\,s$^{-1}$ at $z=0$. J0408-6545 and J1939-6342 were the primary calibrators used for flux density and bandpass calibration, observed at the beginning and end of the observing run. J1311-2216 was observed as a gain calibrator.

\subsubsection{Continuum data reduction}
\label{contiumm_reduction}
The MeerKAT data covering a frequency range of 0.9-1.67~GHz were calibrated using the Containerized Automated Radio Astronomy Calibration \citep[$\tt{CARACal}$;][]{caracal2020} pipeline {\footnote{\url{https://ascl.net/2006.014}}}. The first step consists of flagging in {\tt CARACal}, including shadowed antennas, autocorrelations, and known radio frequency interference (RFI) channels using the tfcrop algorithm.  Thereafter, {\tt AOflagger} \citep{Offringa2010} was used to flag bad data using the {\tt firstpass\_QUV.rfis} strategy. {\tt CARACal} modeled the primary calibrator J0408-6545 using the MeerKAT local sky models. Following this, cross-calibration was performed to solve for the time-dependent delays and complex gains of each antenna and the bandpass corrections.

After initial calibration, we created an initial image of the target field using {\tt WSClean} \citep{Offringa2014} within {\tt CARACal}. Three rounds of phase-only self-calibration were performed using {\tt CubiCal} \citep{Kenyon2018}, followed by a final round of amplitude-phase calibration. The calibrated data were imaged using the Briggs weighting scheme \citep{Briggs1995} with a robust parameter of $0$, and multiscale cleaning.  The properties of the MeerKAT continuum images are summarized in Table\,\ref{con_imaging}.

\subsubsection{Line data reduction}
We only proceed with the data covering the frequency range of 1200-1520\,MHz to limit the computing resources required for calibration and imaging. The data were calibrated on the {\tt ilifu} computer cluster hosted by the Inter-University Institute for Data Intensive Astronomy (IDIA). 

Following \cite{Tirna2021, Tirna_2022}, we first performed the cross-calibration in {\tt CARAcal} (see Section\,\ref{contiumm_reduction} for details). The frequency range covering Galactic emission was flagged. The calibrated measurement set was binned in frequency by a factor of 5 (only for continuum calibration) to achieve a higher signal-to-noise ratio. The continuum imaging was performed in {\tt WSClean} using Briggs weighting with a robust parameter of $-0.5$, and uv taper of 10\arcsec. These parameters were chosen to achieve an optimal balance between angular resolution and sensitivity, with the uv taper specifically applied to recover low surface brightness continuum emission. The {\tt SoFiA} \citep{Serra2015} source finder, integrated into {\tt CARACal}, was utilized to automatically produce a sky mask from the output deconvolved image. The self-calibration solutions were interpolated in frequency and transferred to the cross-calibrated target measurement set. 

We subtracted the sky model using the task {\tt mstransform} within CARACal and fitted a second-order polynomial fit to remove any residual continuum emission from the spectral line data. The data were Doppler corrected to a barycentric reference frame. Finally, spectral line cubes (1370-1425~MHz) were created using WSClean Briggs weighting with a robust value of $-0.5$ and without any uvtaper,  giving an angular resolution of $8.3\arcsec \times 7.6\arcsec$. A number of galaxies in the group have been detected in the new MeerKAT observations. The \hi\ results for other galaxies in the NGC~5044 group will be presented in a later paper. The final \hi cubes were smoothened using binning and boxcart methods, as summarized in Table\,\ref{Tabel:HI_image_properties}.


\begin{figure*}
    \centering
    \includegraphics[width=1.0\textwidth]{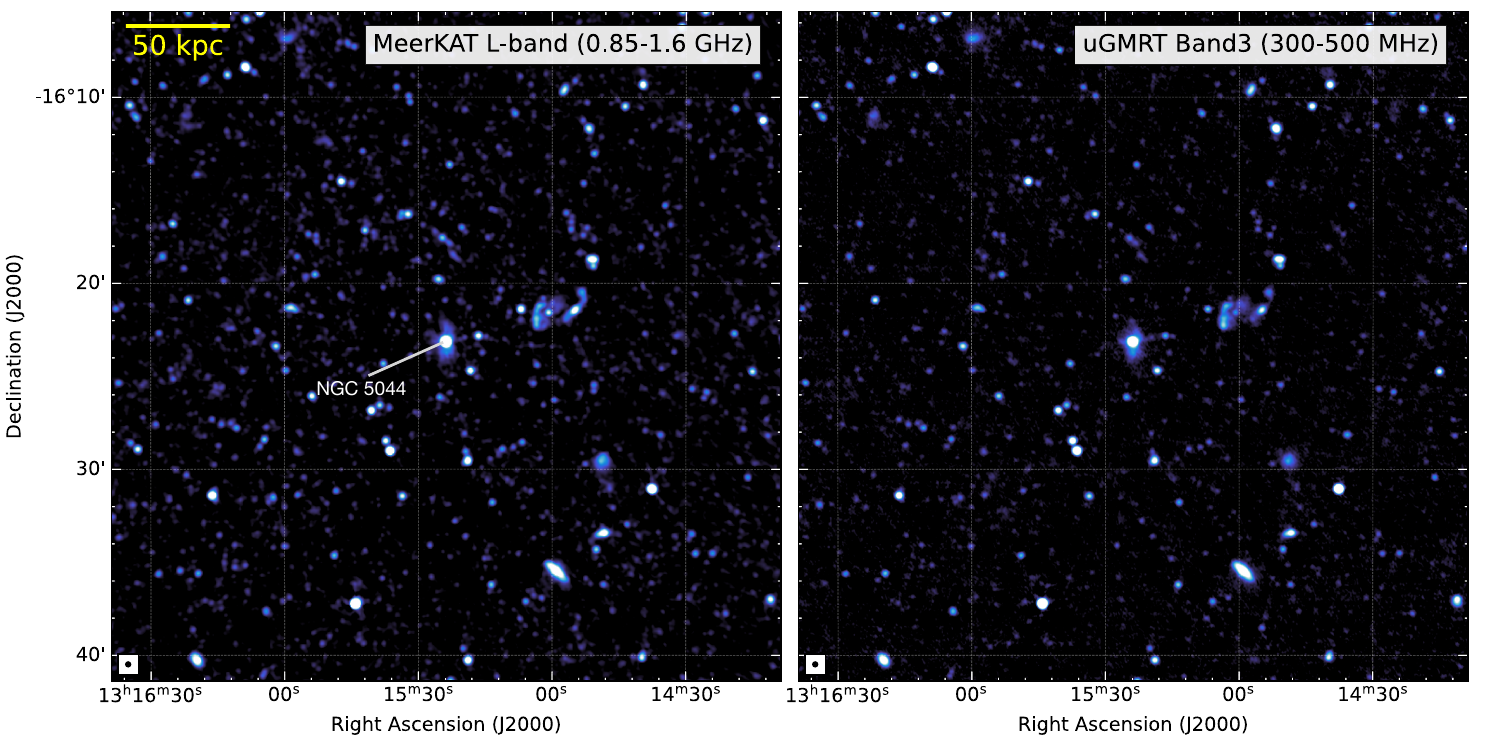}
    \vspace{-0.6cm}
 \caption{MeerKAT L-band (left) and uGMRT Band\,3 (right) full band continuum radio images of the NGC\,5044 field at a common resolution of 15\arcsec. The radio beam size is indicated in the bottom left corner of each image. The image properties are given in Table\,\ref{con_imaging}, listed for IM2 and IM7.} 
      \label{fig1}
\end{figure*}

\setlength{\tabcolsep}{2pt}
\begin{table}
\caption{MeerKAT \hi cubes properties}
\begin{center}
\begin{tabular}{ l  l  l c }
  \hline  \hline  
Frequency range  & $1370-1425$ MHz  \\
Velocity range & $-966.5$  to $11030.1\rm \, km\,s^{-1}$  \\ 
Beam size & $8.3\arcsec \times7.6\arcsec$, $d\nu=5.5\rm \,km\,s^{-1}$ \\
  & ($\rm rms=0.2\,mJy\,beam^{-1}$)  \\
Beam size  & $8.3\arcsec \times7.6\arcsec$, $d\nu=27.9\rm \,km\,s^{-1}$\\
  & ($\rm rms=0.1\,mJy\,beam^{-1}$, binned)  \\  
Beam size & $8.3\arcsec \times7.6\arcsec$ , $d\nu=16.7\rm \,km\,s^{-1}$\\
  & ($\rm rms=0.2\,mJy\,beam^{-1}$, boxcar smoothing)  \\    
\hline 
\end{tabular}
\end{center}
{Notes. All \hi cubes are created with Briggs weighting with ${\tt robust}=-0.5$ and without any uv-cut. All reported rms values represent per-channel noise levels. $d\nu$ is the velocity resolution.}

\label{Tabel:HI_image_properties}
\end{table}


\begin{figure*}
    \centering
    \includegraphics[width=1.0\textwidth]{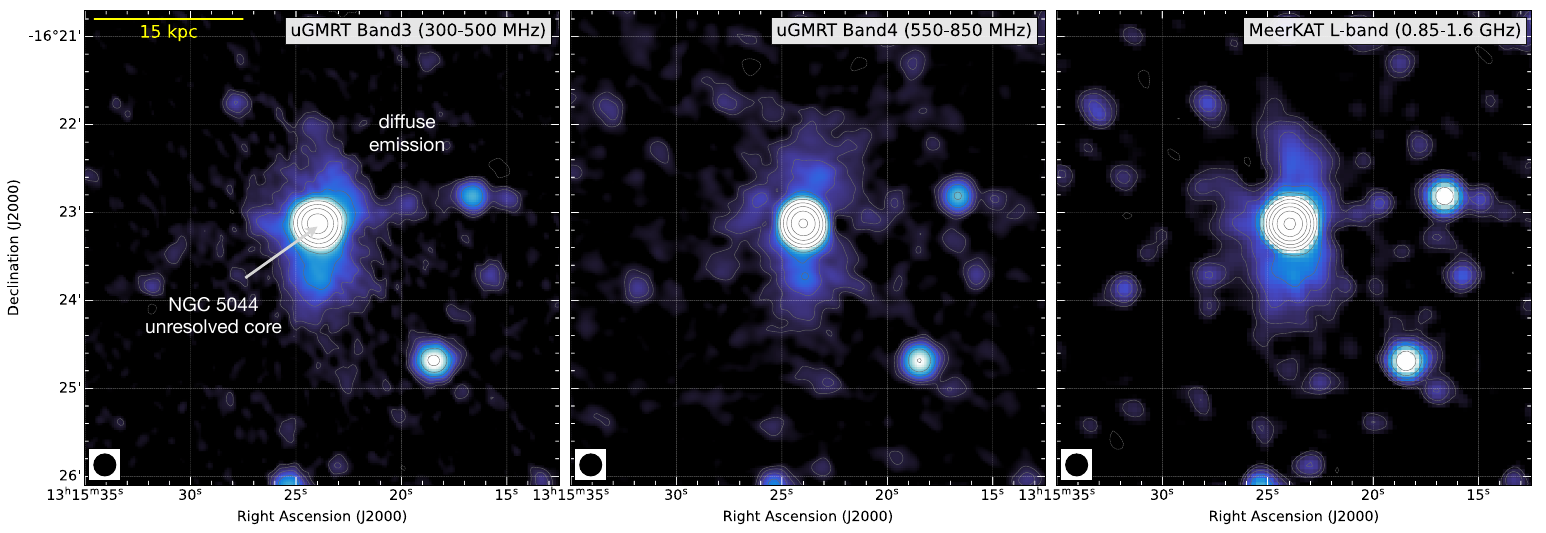}
  \vspace{-0.6cm}  
 \caption{Zoomed-in uGMRT Band\,3 (left), Band\,4 (middle), and MeerKAT L-band (right) continuum images of NGC\,5044, displayed on a square root scale. All images share a common resolution of 15\arcsec. The beam size is indicated in the bottom left corner of each image. The images reveal diffuse radio emission surrounding NGC\,5044 and extending almost 10 kpc from the galaxy in multiple directions. Radio contours are drawn at $[1, 2, 4, 8 ...]\times 3.0\sigma_{\rm rms}$. Dashed contours depict the $-3.0\sigma_{\mathrm{rms}}$ contours. The image properties are given in Table\,\ref{con_imaging}, listed for IM2, IM4, and IM7.}
      \label{zoom_in_NGC5044}
\end{figure*} 


\begin{figure*}
    \centering
             \includegraphics[width=0.458\textwidth]{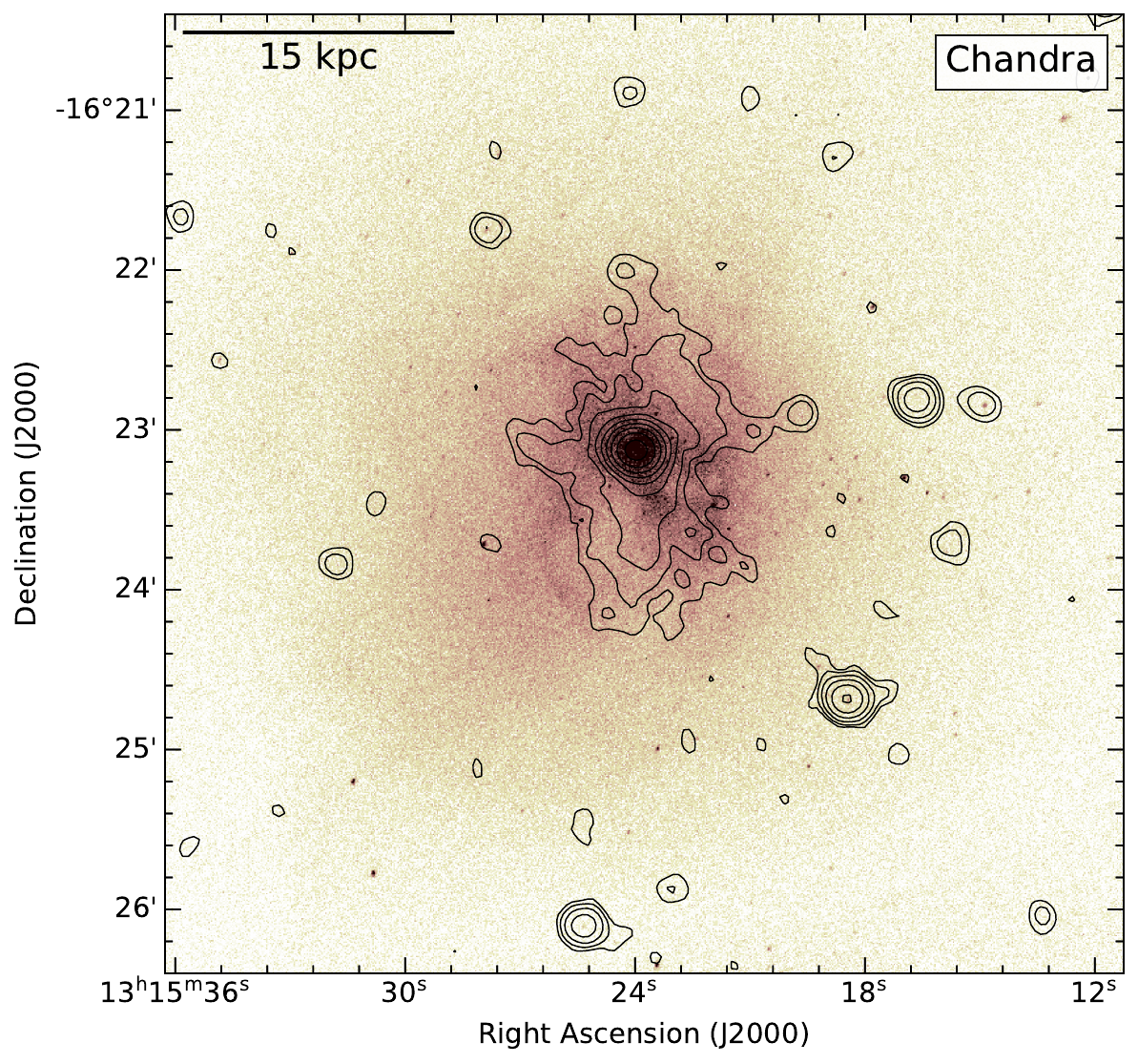}
    \includegraphics[width=0.46\textwidth]{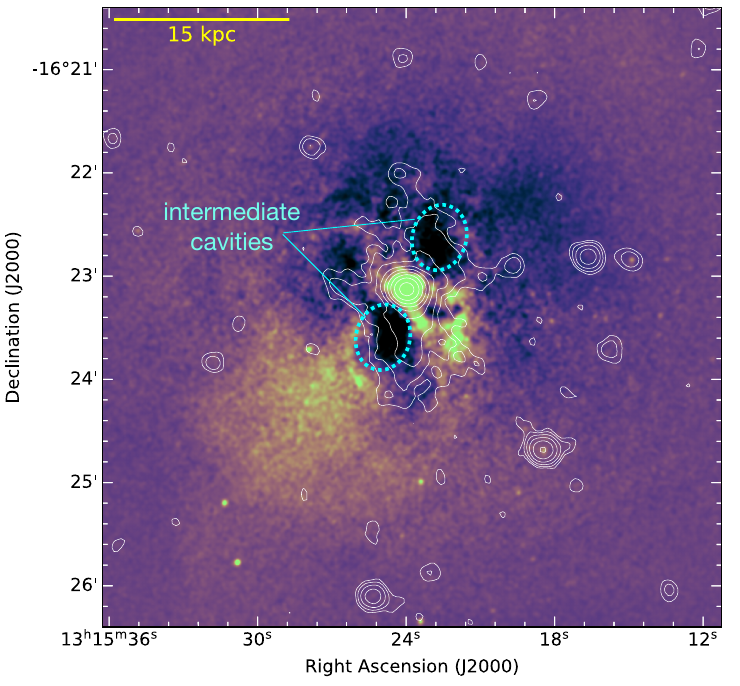}
            \includegraphics[width=0.45\textwidth]{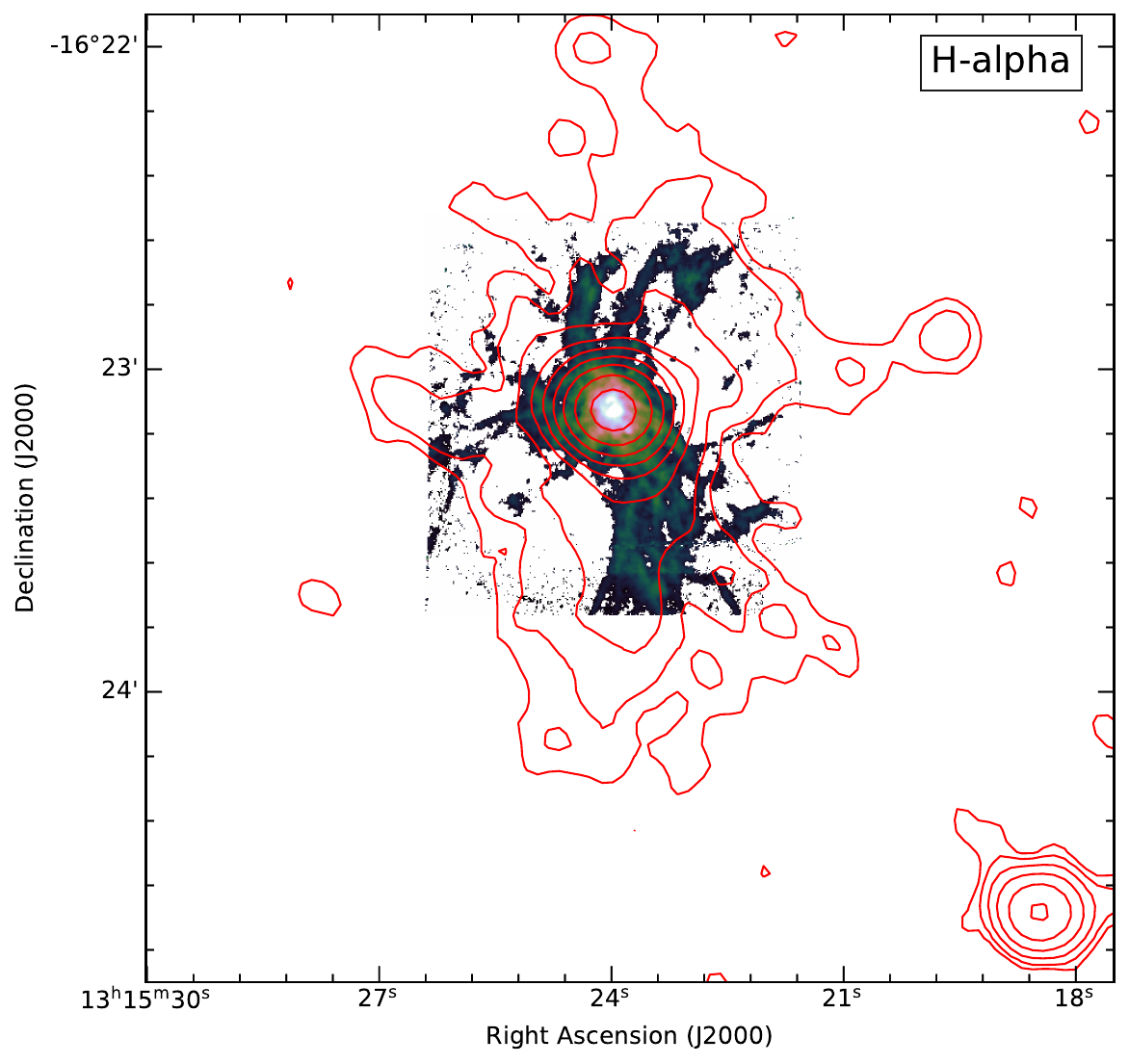}
\vspace{-0.2cm}
\caption{{\it Top Left}: {\chandra} $0.5-2$\,keV image of the galaxy group NGC\,5044 overlaid with the uGMRT Band\,3 10\arcsec\ resolution contours. \textit{Top Right}: {\chandra} $0.5-2$\,keV residual map, smoothed with a 2 pixel ($\sim$1\arcsec) Gaussian and overlaid with the same uGMRT contours. The newly detected diffuse radio emission clearly fills the pair of intermediate cavities and extends beyond. {\it Bottom}: A zoom-in \Ha\ map of the NGC\,5044 overlaid with the uGMRT contours. In contrast to the X-ray, the continuum and \Ha\ are spatially correlated, both are elongated in the north-south direction. In all maps, uGMRT  Band\,3 contour levels are drawn at $[1, 2, 4, 8 ...]\times 3.0\sigma_{\rm rms}$. The radio image properties are given in Table\,\ref{con_imaging}, IM1.} 
      \label{Xray_residual}
\end{figure*}

\subsection{uGMRT}
We observed the group NGC\,5044 with uGMRT in Band\,3 (Project code: 33\_089) and Band\,4 (Project code: 45\_021) using the GMRT Wideband Backend (GWB) and GMRT software Backend (GSB). In both bands, observations were carried out in two observing runs, with 3C147 used as the flux density calibrator. The observational details are summarized in Table\,\ref{Tabel:obs}. 

The uGMRT data calibration was performed using the Source Peeling and Atmospheric Modeling pipeline \citep[\texttt{SPAM}][]{Intema2009}. The main steps are outlined below. Each wideband dataset was split into six sub-bands. The Band\,4 data covering the frequency range 850-950~MHz were heavily affected by radio frequency interference and were thus flagged completely. The flux density of the primary calibrator 3C147 was set according to \cite{Scaife2012}. After flux density scale calibration, the data were averaged, flagged, and corrected for the bandpass. To correct the phase gains of the target field, we started with a global sky model obtained from the GSB data. In the end, the {\tt SPAM} calibrated sub-bands were imaged in {\tt WSClean} using Briggs weighting with robust parameter 0 to produce deep full continuum images. The final deconvolution was always performed using multiscale, see Table\,\ref{con_imaging} for uGMRT continuum image properties.

\section{Results and Analysis }
\label{sec:results}

\subsection{Continuum Emission Properties}
\label{sec::continumm}
In Figure\,\ref{fig1}, we present our deep uGMRT Band\,3 (300-500~MHz) and MeerKAT L-band (0.85-1.67~GHz) images covering a large field of view. A zoom-in of the central region is shown in Figure\,\ref{zoom_in_NGC5044}. These images have a common resolution of 15\arcsec. This resolution allowed us to recover the low surface brightness diffuse emission. Our new observations reveal, for the first time, diffuse radio emission surrounding the central AGN, with a largest linear size of approximately 25~kpc. The continuum emission is elongated along the north-south direction (Figure\,\ref{zoom_in_NGC5044} right panel), in particular toward high frequencies.  Unlike previously published continuum images, the detected radio emission displays a consistent morphology across all observed frequencies, as illustrated in Figure\,\ref{zoom_in_NGC5044}. 

We do not detect any of the previously reported extended structures observed with the GMRT at 235~MHz and 610~MHz or with the VLA at 1.5~GHz \citep{Giacintucci2009, Romana2022}. To investigate further, we reprocessed archival 235\,MHz, 610~MHz, and 150~MHz GMRT observations using the {\tt SPAM} pipeline that includes direction-dependent and ionospheric corrections (see Figure\,\ref{fig:appendix1} and Appendix for data used). The results are consistent with our new uGMRT and MeerKAT maps. We emphasize that our new uGMRT Band\,3 images are a factor of five deeper than the previously published images at 235~MHz. The non-detection of these features in both our new and archival data images suggests that those features are not real. We also reprocessed archival VLA AnB configuration observations. In contrast to \cite{Romana2022}, we do not observe diffuse, jet-like emission extending east-west from the NGC\,5044 core (see Figure\,\ref{fig:appendix2}). Moreover, also in our high-resolution MeerKAT L-band image, we find no indication of the east-west jet-like emission (see Figure\,\ref{fig:appendix2}).

\begin{figure*}
    \centering
    \includegraphics[width=0.45\textwidth]{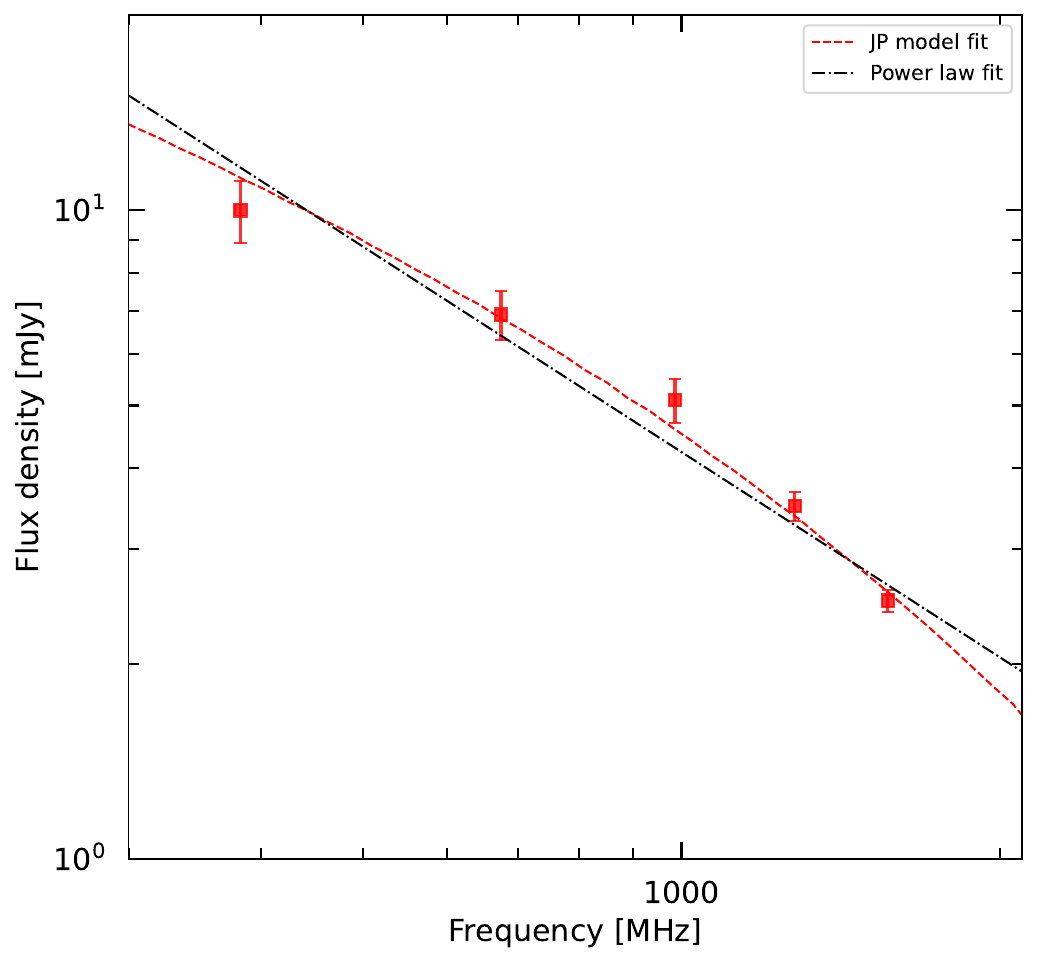}
    \includegraphics[width=0.46\textwidth]{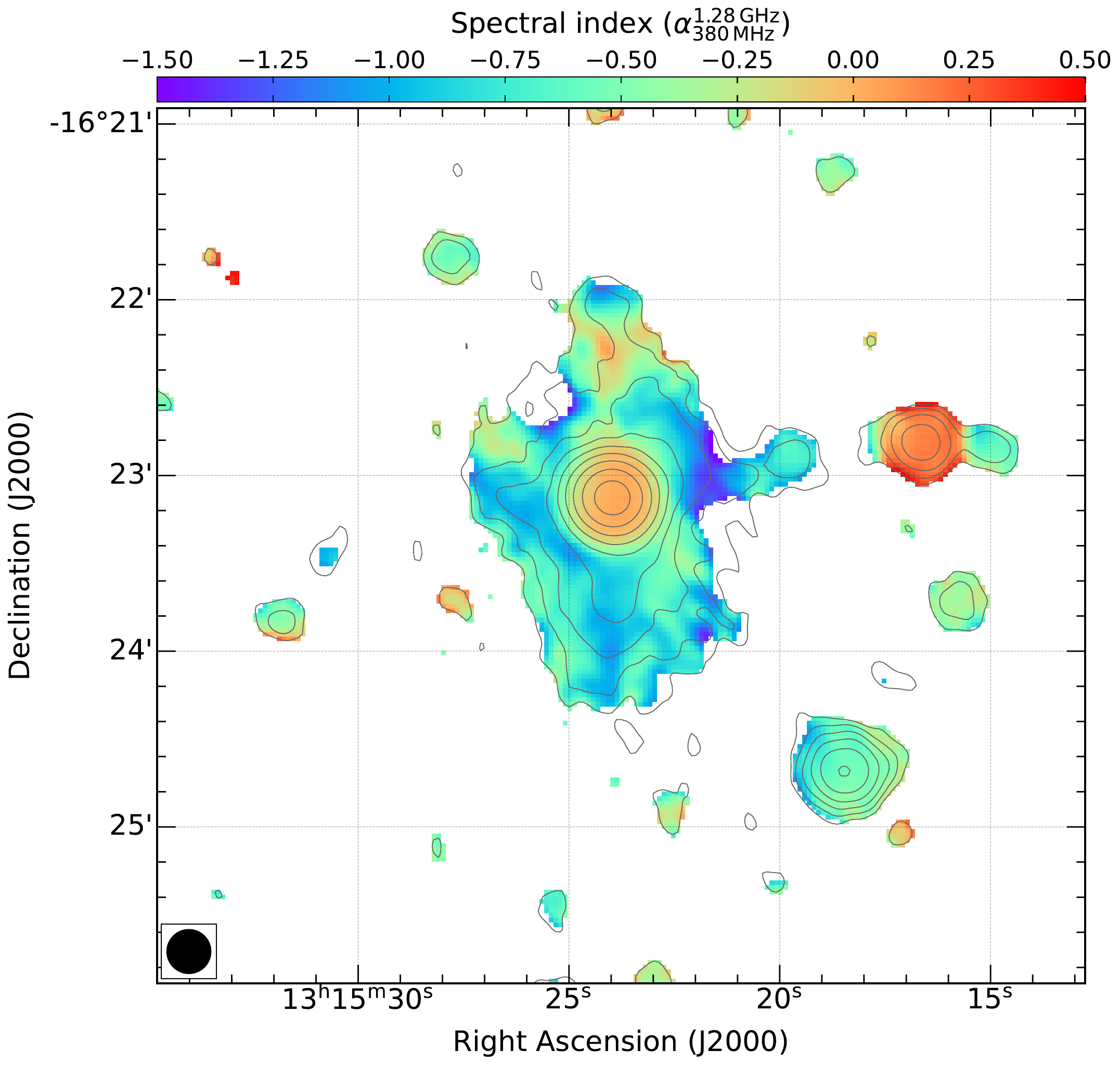}
 \caption{\textit{Left}: Integrated spectrum of the newly detected diffuse radio emission excluding NGC\,5044 core (i.e., within a 6.5\,kpc diameter) between 380~MHz and 1.56~GHz. The dashed and dot-dashed lines represent the fitted JP model and a single power law, respectively. The JP model fitting is performed in {\tt synchrofit}. \textit{Right}: Spectral index map at 15\arcsec resolution created between 380~MHz and 1.28~GHz, showing no particular trends. Radio contour levels are drawn at $[1, 2, 4, 8 ...]\times 3.0\sigma_{\rm rms}$. The image properties are given in Table\,\ref{con_imaging}, listed for IM3 and IM8.}
     \label{radio_index}
\end{figure*}

In Figure\,\ref{Xray_residual} (top-left panel), we compare the radio and X-ray morphologies. A {\chandra} $0.5-2$~keV map is overlaid with the uGMRT 380\,MHz 10\arcsec\ resolution contours. We used a higher resolution (10\arcsec) radio image, as it resolves certain radio structures and allows for a better comparison with X-ray features. Overall, the morphology of the newly detected radio emission does not correlate with the X-ray emission. Several structures are seen in the X-ray image with no clear radio counterpart. Moreover, on large scales (i.e., intermediate cavities and regions beyond them) there is only a partial correlation between radio and X-ray structures. In the top-right panel of Figure\,\ref{Xray_residual}, we show a {\chandra} $0.5-2$~keV residual map \citep[obtained by subtracting the best-fitting elliptical $\beta$-model, convolved with the exposure map, from the merged image,][]{David2017}. Clearly, the newly detected radio emission extends significantly beyond the intermediate cavities (dashed cyan regions). The radio emission is moderately brighter in the northern intermediate cavity while it appears fainter (i.e., about a factor of 2 less brighter) and less prominent in the southern cavity. Although the cavities are fully filled by the radio emission, they (cavities) do not exhibit a clear connection to the overall radio morphology. In contrast to the radio and X-ray emission, there seems to be a spatial coincidence between the \Ha\ and the radio emission, as both are elongated in the north-south direction (see bottom panel of Figure\,\ref{Xray_residual}).

\begin{table}
\caption{Flux densities of the diffuse emission excluding the NGC\,5044 core (i,e., within a 6.5\,kpc diameter)}
\begin{center}
\begin{tabular}{ c  c  c c c c}
  \hline  \hline  
\multirow{1}{*}{Frequency} & \multirow{1}{*}{}{ Flux density} \\ 
\multirow{1}{*}{[MHz]} & \multirow{1}{*}{[mJy]} \\
\hline
380 & $10\pm1.0$  \\
675 & $6.9\pm0.6$  \\
986 &  $5.1\pm0.5$ \\
1228 & $3.5\pm0.1$  \\
1567& $2.5\pm0.1$ \\
\hline 
\end{tabular}
\end{center}
{Notes. Flux densities were measured from 15\arcsec~resolution images, created with Briggs weighting with $\rm {\tt robust}=0$ and an common uv-cut of $200\,k\lambda$ (see Table\,\ref{con_imaging}, IM3, IM5, and IM8.). We assumed an absolute flux scale uncertainty of 10\% for the uGMRT and MeerKAT UHF-band, and 5\% for L-band.}
\label{Tabel:flux}
\end{table} 

To investigate the overall radio properties of the newly detected radio emission, we performed integrated and spatial spectral analyses. For the integrated radio analysis, we used wideband 15\arcsec\ images at 380 MHz, 675 MHz, and 1.28 GHz, created with a uv-cut of $200\,k\lambda$ and Briggs weighting with ${\tt robust}=0$. The same images were used to produce a resolved spectral index map. To constrain the spectrum better, especially to check for a possible spectral break toward higher frequencies, we also include MeerKAT 986~MHz and 1.56~GHz sub-band images. The measured flux densities of the diffuse radio emission (i.e., excluding the contribution from the NGC 5044 core) are summarized in Table\,\ref{Tabel:flux}. The resulting integrated spectrum is shown in Figure\,\ref{radio_index} left panel, revealing a high frequency spectral steepening. At low frequencies, the spectral index is $\alpha_{\rm 380\,MHz}^{986\,\rm MHz} = -0.70\pm0.05$, which steepens to $\alpha_{986\,\rm MHz}^{1.56\,\rm GHz} = -1.53\pm0.06$ at higher frequencies. The trends appear consistent with aging of electrons from AGN lobes owing to synchrotron and inverse Compton losses.

In Figure\,\ref{radio_index} right panel, we present the spectral index map of the central region of NGC\,5044, created using maps between 380~MHz and 1.28~GHz. Pixels with flux density below $3\sigma_{\rm rms}$ were blanked. We do not observe any specific trends across the diffuse emission, i.e., no clear evidence of spectral index gradients in a particular direction. The spectral age of the detected diffuse emission is determined by fitting the integrated spectrum of two subregions (only included diffuse emission) with a single injection Jaffe-Perola model \citep[JP;][]{Jaffe1973}. We emphasize that the compact core region of the NGC\,5044 is excluded from the age analysis. Following \cite{Brunetti1997}, for a power law energy distribution $N(E)\propto E^{-p}$, we estimated the magnetic field using the minimum-energy assumption:

\begin{equation}
\begin{aligned} {B_{\rm eq,rev}=D(p)\gamma _{\min }^{\frac{2(2-p)}{p+5}}B_{\rm eq},^{\frac{7}{p+5}}} ,\end{aligned} 
\end{equation}
where $B_{\mathrm{eq}}$ is as follows:

\begin{equation}
\begin{aligned}
B_{\mathrm{eq}} = 5.69 \times 10^{-5} \left[ \frac{(1 + k)}{\eta} (1 + z)^{3 - \alpha} 
\frac{1}{\theta_x \theta_y \, l \sin^{3/2} \phi} \right. \\
\times \left. \frac{S_{\rm obs}}{\nu_{\rm obs}^{+\alpha}} 
\frac{\nu_2^{0.5 + \alpha} - \nu_1^{0.5 + \alpha}}{0.5 + \alpha} \right]^{2 / 7} \, \text{Gauss}.
\end{aligned}
\end{equation}
$\theta_x$ and $\theta_y$ represent the angular dimensions of the radio source in arcseconds along the x- and y-axes, respectively. $k$ is the proton-to-electron energy ratio, $\eta$ is the filling factor of the emitting region, and $l$ denotes the path length through the source in kiloparsecs. $S_{\text{obs}}$ is the radio flux density (in Jy) of the region at the observed frequency $\nu_{\text{obs}}$ (in GHz). $\nu_1$ and $\nu_2$ are the pre-assumed lower and upper cut-off frequencies for the radio spectrum (in GHz). $\phi$ is the angle between the magnetic field and the line of sight. The particle energy distribution index, $p$, is given by $p = 1-2\alpha$, and $\gamma_{\text{min}}$ represents the minimum Lorentz factor of the particle distribution.

We use a spectral index of $-0.55$ (i.e., injection index obtained by fitting the spectrum with the JP model using {\tt synchrofit}{\footnote{\url{https://github.com/synchrofit/synchrofit}}}), $\nu_1=0.010~\rm GHz$, $\nu_2=100~\rm GHz$, $S_{obs}=9.9\,\rm mJy$, $\nu_{obs}=0.380~\rm GHz$, and assume $k=1$, $\eta=1$. For size, we adopted $\theta_x=66\arcsec$ and $\theta_y=110\arcsec$, and the radial direction $l=9.9\rm\,kpc$ (assuming $l$ is equal to the transverse direction), all measured from the uGMRT Band\,3 map, and excluded the NGC\,5044 core region. Following \cite{Brunetti1997}, we adopt  $D(\gamma) = 1.01$  and $\gamma_{\text{min}} = 20$. This gives us the revised equipartition magnetic field value of $B_{\rm eq,rev}= 1.4\rm \mu G$, resulting in an estimated mean spectral age of $77\pm10$ Myr. Additionally, we also consider a magnetic field value that minimizes the radiative losses and maximizes the lifetime of the source ($B=B_{\rm CMB}/\sqrt{3}$ where $B_{\rm CMB}=3.25(1+z)^2$ is the equivalent magnetic field of the cosmic microwave background), estimated to be approximately $1.9 \rm \mu G$. This yields a mean spectral age of about $80\pm15$ Myr.

\begin{figure}
    \centering
        \includegraphics[width=0.46\textwidth]{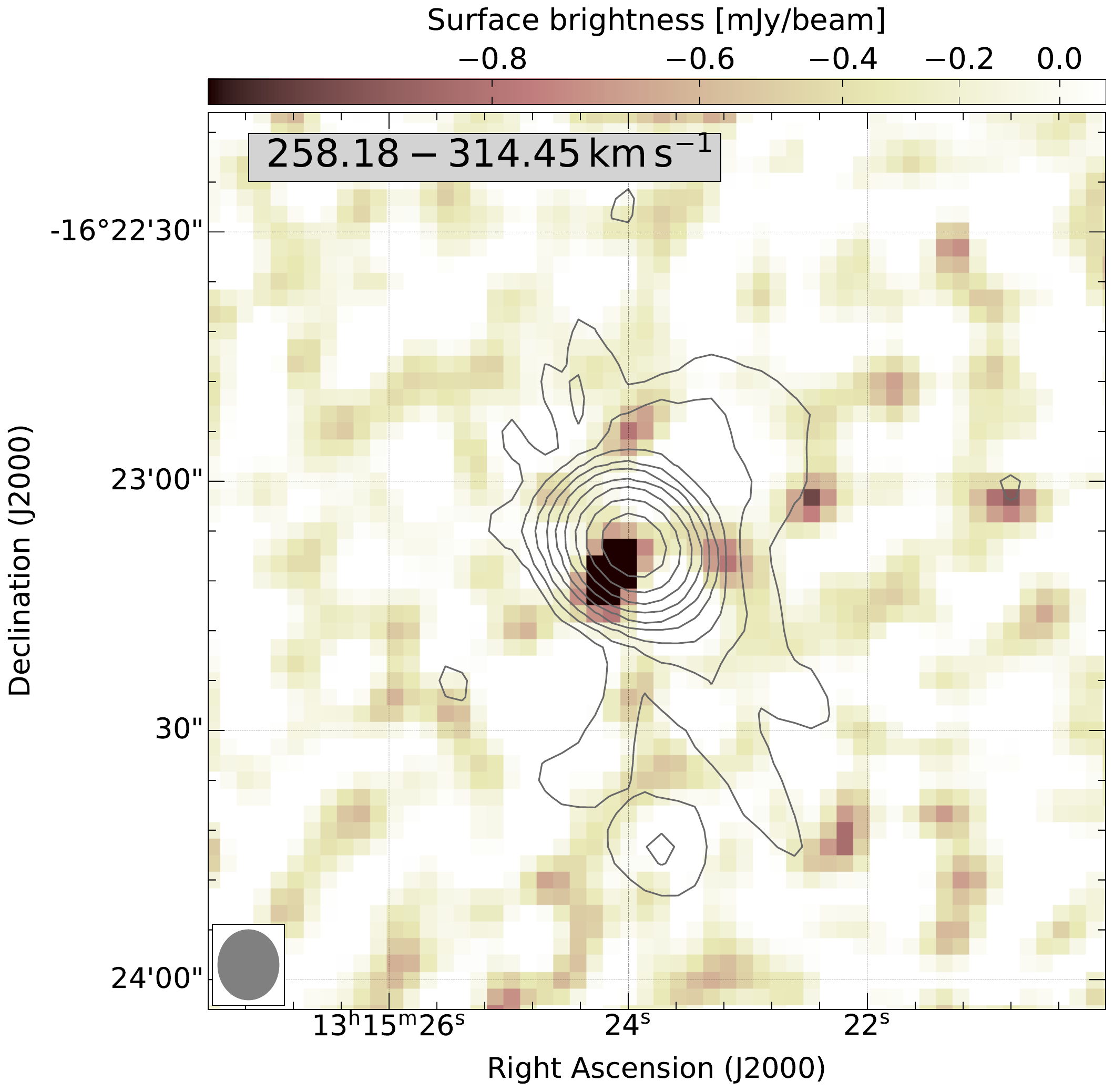}
 \caption{\hi moment zero MeerKAT map ($8.3\arcsec \times 7.6\arcsec$) showing absorption from the core of NGC\,5044 overlaid with continuum emission contours at 1.28 GHz ($8.3\arcsec \times 7.6\arcsec$). The \hi\, cube was binned to a velocity resolution of $\rm 27.9\,km\,s^{-1}$ (5 channel binning). The image reveals the \hi absorption from the core region of NGC\,5044. Contour drawn at $[1, 2, 4, 8 ...]\times 5.0\sigma_{\rm rms}$ where $\rm rms=12\,\rm \mu Jy\,beam^{-1}$.}
     \label{HI_abs}
\end{figure} 


\begin{figure*}
    \centering
    \includegraphics[width=0.488\textwidth]{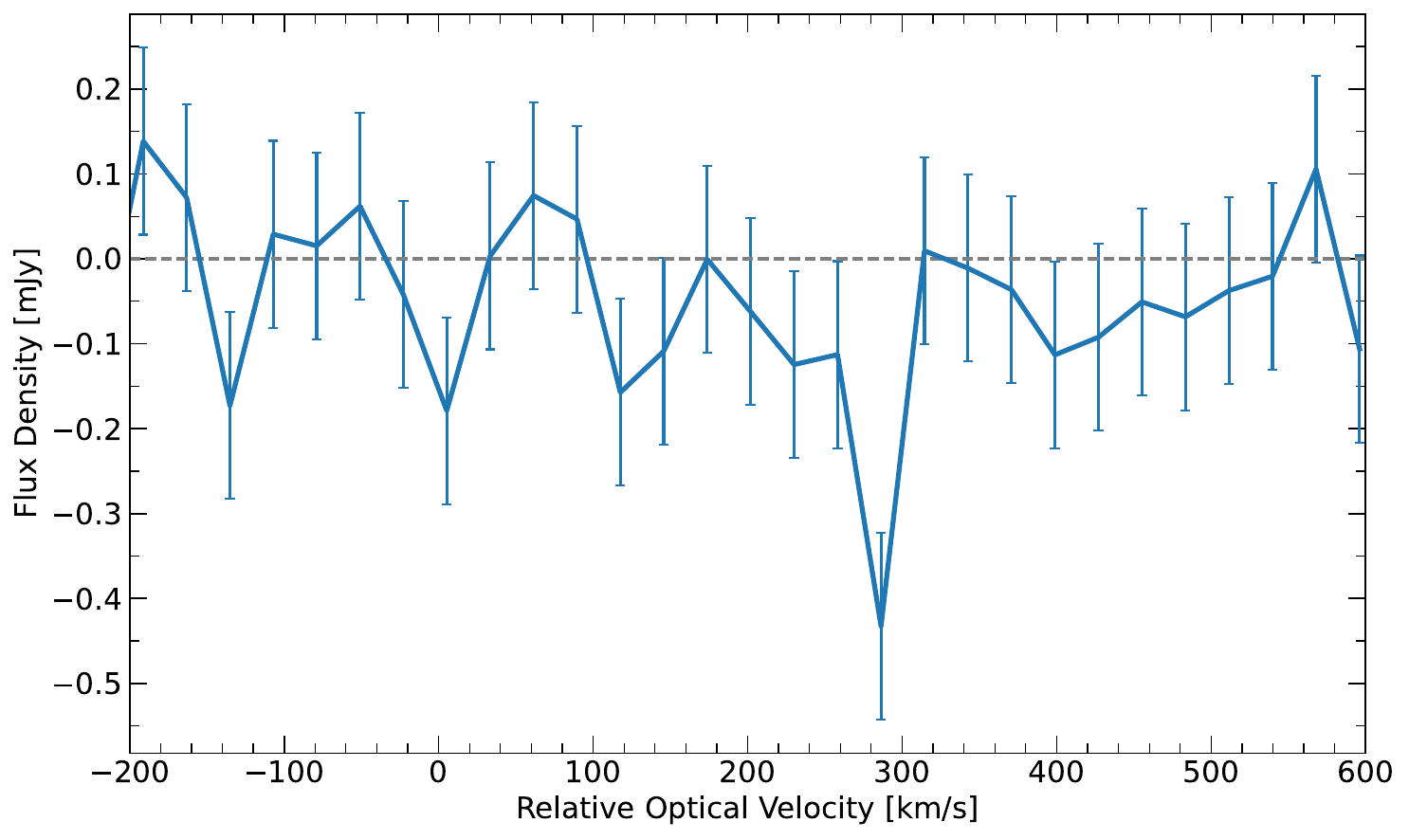}
      \includegraphics[width=0.49\textwidth]{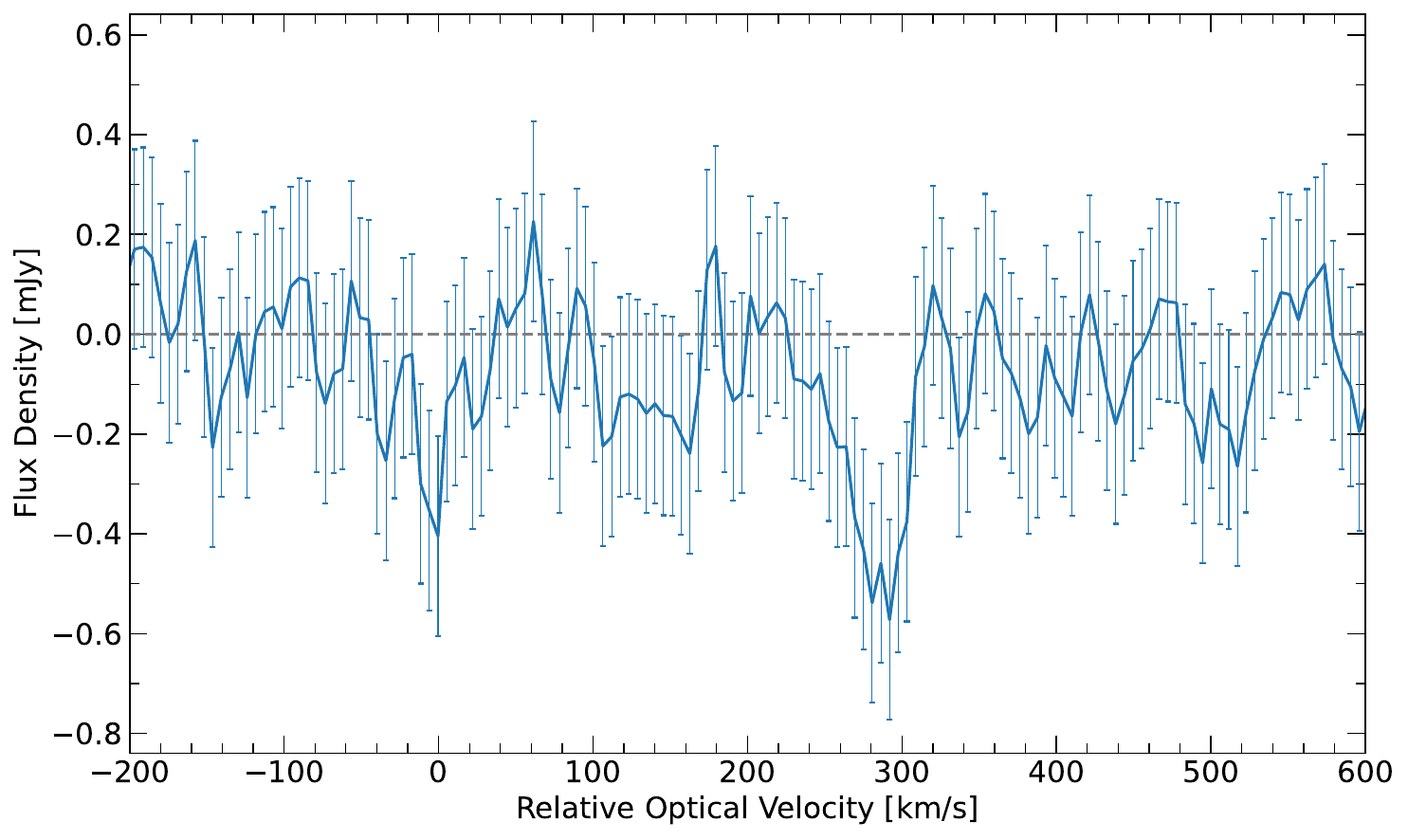}
 \caption{ \hi absorption spectra toward the compact core of NGC\,5044 obtained from {\tt SoFiA}. Both spectra are extracted from an \hi cube with beam size of $8.3\arcsec \times 7.6\arcsec$ using a circular region of radius 4.2\arcsec, i.e, similar to the beam size. \textit{Left}: Spectrum obtained by applying 5-channel binning, resulting in an effective velocity resolution of $\mathrm{27.9 \, km \, s^{-1}}$. The rms noise of the \hi cube is $\mathrm{0.1 \, mJy \, beam^{-1}}$. \textit{Right}: The spectrum was obtained by performing Boxcar smoothing with a 3-channel kernel, resulting in an effective velocity resolution of  $\rm 16.7\,km\,s^{-1}$.}
      \label{HI_spectra}
\end{figure*} 


\subsection{H\textsc{i} Properties}
In Figure\,\ref{HI_abs}, we show our \hi moment zero map (i.e., integrated intensity) from the MeerKAT \hi cube centered on the core which reveals \hi absorption detected towards NGC\,5044. The \hi cube was binned to a velocity resolution of $\rm 27.9\,km\,s^{-1}$ (i.e., 5 channel binning) to enhance the sensitivity of the data. The \hi absorption is concentrated within the core region, i.e., within the central $~800$~pc (5.4\arcsec) diameter.

\begin{figure*}
    \centering
    \includegraphics[width=0.7\textwidth]{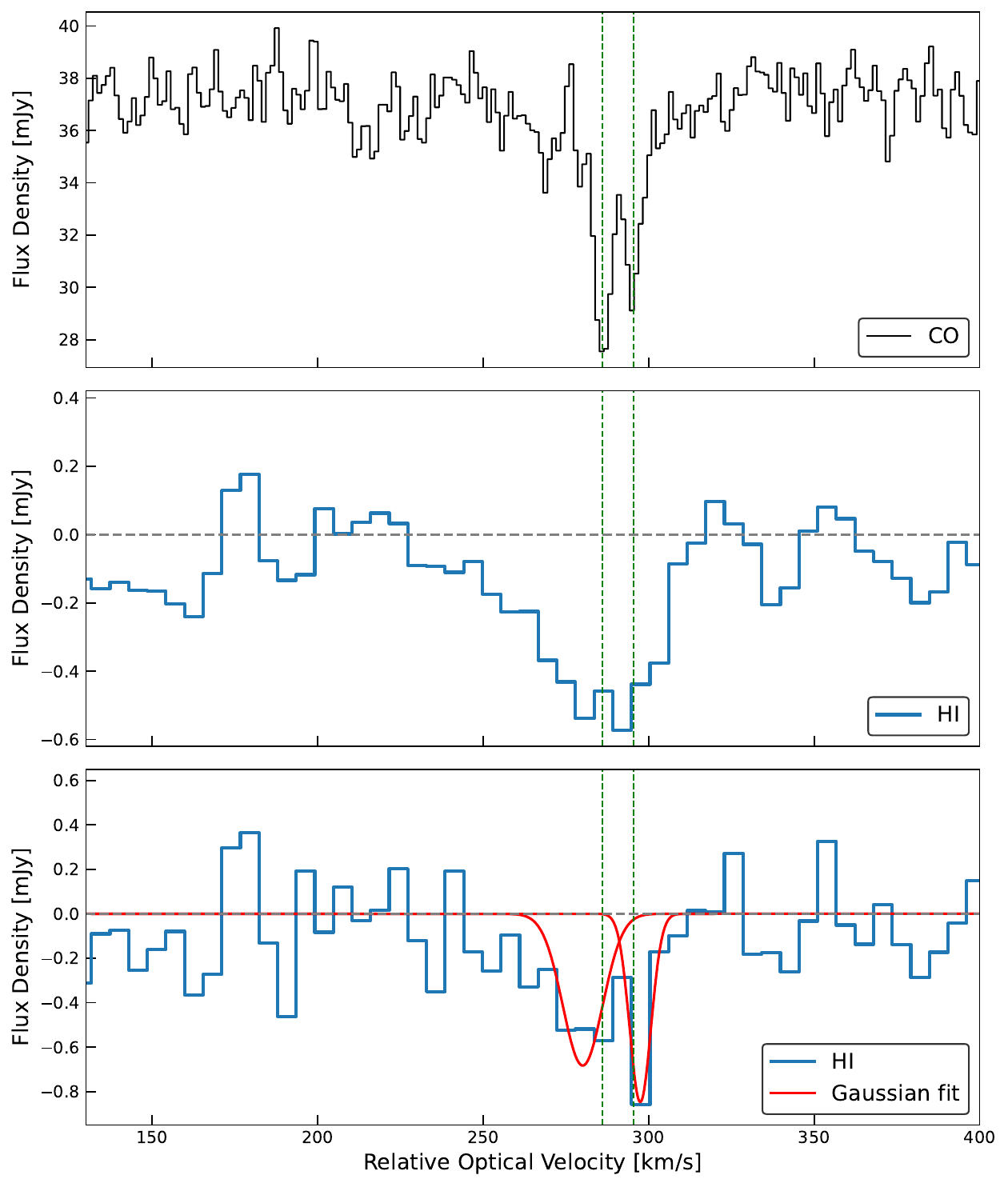}
 \caption{CO (top) and \hi (middle/bottom) absorption spectra from the nuclear region of NGC~5044. The raw \hi spectrum is shown in the lower panel, and the middle panel shows the spectrum, boxcar smoothed to a velocity resolution of $\rm 16.7\,km\,s^{-1}$. The CO spectrum, adopted from \cite{Schellenberger2020}, shows two strong absorption features peaks at $\nu=\rm 286.1\,km\,s^{-1}$ and $295.5 \rm\,km\,s^{-1}$ (green dashed lines). The \hi absorption also shows a double-dip and exhibits a broader tail toward lower velocities compared to the CO. The two-Gaussian \hi components are shown in red (bottom) and the values are given in Table\,\ref{Tabel:double_gaussian_fititng}. \hi spectra (middle and bottom) are extracted within 4.2\arcsec\, radius. }
      \label{HI__CO_spectra}
\end{figure*} 

Figure\,\ref{HI_spectra} (left panel) shows the resulting \hi spectrum, generated in {\tt SoFiA} by integrating over a 2D mask (a circle with a radius of 4.2\arcsec, which is similar to the beam size) centered on NGC\,5044 in each channel. The spectrum shows a pronounced absorption feature at $\rm 286\,km\,s^{-1}$. The detection has a significance of $3.8\sigma$ (99.9\% confidence). 

We also applied Boxcar smoothing to assess the robustness of the detected \hi absorption feature. Using a 3-channel kernel, this smoothing resulted in a velocity resolution of $\rm 16.7\,km\,s^{-1}$. This resolution was specifically chosen to investigate the potential presence of double absorption features, as seen in the CO observations. The corresponding spectrum is shown in Figure\,\ref{HI_abs} right panel and reveals a double absorption dip structure. We note at the same velocity resolution, the binning method also reveal the double-dip absorption features, further affirming the presence of double-peaked absorption.


\begin{table}
\caption{Gaussian fits to the \hi and CO absorption features}
\begin{center}
\begin{tabular}{ c  c  c c c c}
  \hline  \hline  
& \multirow{1}{*}{Velocity} & \multirow{1}{*}{}{ FWHM} & \multirow{1}{*}{}{Amplitude} \\ 
&\multirow{1}{*}{($\rm km\,s^{-1}$)} & \multirow{1}{*}{($\rm km\,s^{-1}$)} & \multirow{1}{*}{(mJy)}\\
\hline
\hi & $280.1\pm2.1$ & $14.2\pm5.4$ &$0.68\pm0.19$ \\
& $297.5\pm1.7$ & $7.6\pm2.3$ & $0.85\pm0.2$\\
\hline 
CO & $286.1\pm0.3$ & $6.7\pm0.7$ &$9.5\pm0.7$ \\
& $295.5\pm0.4$ & $6.4\pm0.9$ &$7.2\pm0.7$ \\
\hline 
\end{tabular}
\end{center}
{Notes. Double Gaussian fit for the \hi spectrum is extracted within an aperture representing the beam size. The cube has a beam size of $8.3\arcsec \times 7.6\arcsec$, velocity resolution $5.5\,\rm km\,s^{-1}$, and rms $0.2\,\rm mJy\,beam^{-1}$. We emphasize that the CO spectrum is adopted from \cite{Schellenberger2021} but using the optical velocity definition.}
\label{Tabel:double_gaussian_fititng}
\end{table} 

\subsubsection{H\textsc{i} column density}

The column density of the \hi absorption feature is estimated as: 
\begin{equation}
\frac{N_{\rm \hi,\,abs}}{{\rm cm^{-2}}} = 1.82 \times 10^{18}  \, \left(\frac{T_{\text{spin}}}{K}\right) \int \tau \, \left( \frac{dv}{\rm km\,s^{-1}} \right),
\end{equation}
where $T_{\rm spin}$ is the spin temperature representing the weighted harmonic mean of the various thermal components of \hi. The $ \int \tau \, dv$ is the optical depth of the \hi absorption calculated as: 
\begin{equation}
\int \tau \, dv = \sum_{i=1}^{n} -\ln\left(1 - \frac{f_i}{S}\right) \times \Delta v,
\label{equ:tau}
\end{equation}
where ${f_i}$ is the \hi absorption flux density in Jy at frequency channel $i$, $n$ is the number of channels where absorption is seen, $S$ is the continuum flux density in Jy, and  $\Delta v$ is the width of each channel in $\rm km\,s^{-1}$. We measure a continuum flux density of $31\pm2$ mJy (extracted from the full wideband image) at 1.4~GHz. Substituting $\Delta v =5.5\rm km\,s^{-1}$ and other values in Equation\,\ref{equ:tau}, we calculate $\int \tau \, dv =0.526\pm0.103\,\rm km\,s^{-1}$. This optical depth corresponds to an \hi absorption column density of $N_{\rm \hi,\,abs}=9.58\pm1.87 \times 10^{17} \, T_{\rm spin}\,\text{cm}^{-2}$. To accurately constrain the \hi column density, it is crucial to estimate the spin temperature. For instance, a dense cold neutral medium is characterized by spin temperatures in the range $100 \leq T_{\rm spin} \leq 1000\,\text{K}$ \citep{Wolfire1995}, whereas the warm neutral medium exhibits significantly higher spin temperatures, ranging from $5000 \leq T_{\rm spin} \leq 8000\,\text{K}$. 

The spin temperature can be directly estimated from the ratio of the emission and absorption column densities, where the \hi emission column density is given by \citep{Allison2021}:
\begin{equation}
N_{\text{\rm \hi,\,em}} = 1.823 \times 10^{18} \int T_B \, dv,
\end{equation}
where $T_B$ is the brightness temperature in units of $K$. The above equation is derived under the assumption that the background radiation is negligible and approximating  $T_{\rm spin} \int \tau_{\hi} \, dv \approx \int T_B \, dv$ \citep[for details see][]{Meyer2017, Allison2021}. The velocity integrated $T_B$ can be calculated as : 
\begin{equation}
\int T_B \, dv = 1.36 \times \lambda^2 \times \frac{1}{\theta_a \, \theta_b} \, S_{\text{int}},
\label{equ:Tspin}
\end{equation}
where $S_{\text{int}}=\int S_{\nu} \, dv$ is the integrated \hi emission flux density in $\rm mJy\,km\,s^{-1}$, $\theta_a$ and $\theta_b$  are the full width at half-maximum (FWHM) of the synthesized beam in arcsec. However, as the 21 cm line was not detected in emission, we can only put an upper limit on the spin temperature. The $3\sigma$ upper limit for the integrated line flux density is $S_{\text{int}}\leq 0.0522\,\rm Jy \,km\,s^{-1}$ (see the next section).  Putting this $S_{\text{int}}$ value, $\theta_a=8.3\arcsec$, $\theta_b=7.6\arcsec$, and $\lambda=21.1\rm cm$ (for \hi line) in Equation\,\ref{equ:Tspin}, we obtain the upper limits $T_B\leq 500 K$ and $N_{\text{\rm \hi,\,em}}\leq 9.13\times10^{20}\,\rm cm^{-2}$. By dividing this column density by our estimated absorption column density, we get the spin temperature upper limit of $T_{\rm spin}\leq 950\,\rm K$, suggesting that the detected \hi is in the cold neutral
phase. \cite{Schellenberger2020} obtained a molecular hydrogen column density of $N_{H_2} \approx 1.5 \times 10^{20} \, \text{cm}^{-2}$, from the CO absorption.

\subsubsection{H\textsc{i} mass upper limit}
In our MeerKAT L-band observation, no \hi emission is detected toward NGC\,5044. We place an upper limit on the \hi gas mass from the non-detection:
\begin{equation}
\frac{M_{\hi}}{M_{\odot}} =  \frac{2.36 \times 10^5 \,{D_L^2}}{(1+z)} \int S_{\nu} \, dv,
\label{eq: hi_mass}
\end{equation}
where $  \int S_{\nu} \, dv$ is the integrated line flux density in $\rm Jy \,km\,s^{-1}$ and $D_{\rm L}$ is the luminosity distance to the galaxy in Mpc.  We derived a $3\sigma$ upper limit for the integrated flux density $ \int S_{\nu}\, dv=3\sigma_{\rm rms}\sqrt{N}\Delta v$, where N is the number of channels that the galaxy would span, $\sigma$ is the noise per channel and $\Delta v$ is the channel width. Following  \cite{Schellenberger2020}, we determined N using the CO(2-1) emission velocity range which spans from $-520$ and $330\,\rm km\,s^{-1}$. This results in $ \int S_{\nu}\, dv< 0.0522\,\rm Jy \,km\,s^{-1}$ (within 4.2\arcsec\ radius aperture). Since the CO integrated flux density was estimated within a 15\arcsec\, radius aperture, we also apply the same aperture, yielding $ \int S_{\nu}\, dv< 0.2367\,\rm Jy \,km\,s^{-1}$. Substituting $\rm D_{L}=31.2\rm\,Mpc$  and other relevant parameters into Equation\,\ref{eq: hi_mass},  we obtained an \hi mass upper limit of $M_{\hi}< 1 \times10^7 M_{\odot}$ (4.2\arcsec\, radius aperture) and $M_{\hi}< 5.4 \times10^7 M_{\odot}$ (15\arcsec\, radius aperture). 

\cite{Schellenberger2020} reported a molecular mass of $M_{\rm H_2}=9.5\pm1.2 \times10^7 M_{\odot}$ in a 15\arcs\ radius aperture using the standard Milky Way CO-to-H2 conversion factor of $X_{\rm CO}=2\times 10^{20}\,\rm  cm^{-2} (K\,km\,s^{-1})$. This implies that the ratio of molecular to atomic gas in the central 2.2~kpc of NGC~5044 is $>$1.76:1. We note that it is unclear whether the Galactic conversion factor is applicable in the cooling cores of galaxy clusters and groups. As yet only one measurement is available, which finds a value a factor of 2 lower \citep{Vantyghem2017}, but more measurements are needed to determine whether this is representative of the population as a whole. In NGC~5044, we can only note that the molecular gas mass could be a factor of 2 lower, in which case our \hi mass upper limit could be comparable to the measured H$_2$ mass.

\subsubsection{H\textsc{i} and CO(2-1) comparison}

In Figure\,\ref{HI__CO_spectra}, we compare the \hi spectrum with the CO(2-1). Clearly, the \hi absorption feature has a counterpart in CO(2-1) and is moderately  broader. We fit the \hi absorption features with a double Gaussian model. For fitting we extracted the spectrum from the unsmoothed cube (i.e., $5.5\,\rm km\,s^{-1}$ velocity resolution) which is essential for accurately determining the intrinsic shapes and widths of the spectral features. The resulting fit is shown in the bottom panel of Figure\,\ref{HI__CO_spectra}. The fitting yields velocities of $\rm 280.1\pm2.1\,km\,s^{-1}$ and $\rm 297.5\pm1.7\,km\,s^{-1}$, comparable to the CO(2-1) velocity values. In  Table\,\ref{Tabel:double_gaussian_fititng} we present the FWHM and velocities of each Gaussian component.

Comparing the $27.9\,\rm km\,s^{-1}$ smoothed \hi spectrum (Figure\,\ref{HI_spectra} left panel) reveals that its peak velocity aligns closely with one of the two CO(2-1) absorption peaks at $\rm 286.1\, km\,s^{-1}$. The higher-resolution $5.5\,\rm km\,s^{-1}$ \hi spectrum (Figure\,\ref{HI__CO_spectra}) reveals a double-dip and one of them approximately corresponds to the velocity of the CO absorption features (indicated by green dashed lines; see also Table\,\ref{Tabel:double_gaussian_fititng}). 


\section{Discussion}

\subsection{Origin of the diffuse radio emission}
As noted in Section~\ref{sec:intro}, the radio structure of NGC~5044 has been a subject of some debate, with only the VLBA observations of the parsec-scale core and jets providing a consistent picture at multiple frequencies. On scales larger than a kiloparsec, prior published observations at different frequencies presented different radio structures, with only partial correlations with the cavities and filaments observed in the hot, X-ray emitting IGrM \citep{Giacintucci2011,David2011,O'Sullivan2014,Romana2022}. Several factors may have contributed to this lack of consistency between the published radio images, and between them and our new analysis: 1) the relative faintness of the diffuse emission in NGC~5044, only $\sim$1\mJypb\ at L-band for the brightest areas; 2) the relatively shallow older archival VLA and GMRT observations, made prior to broad-band upgrades which dramatically improved the uv-coverage and sensitivity of these telescopes; 3) the advent of, and improvements in, tools for ionospheric modeling and direction-dependent calibration; 4) Advancements in data reduction pipelines and particularly new image reconstruction algorithms (that use, for instance, the multiscale option, which is crucial to properly image low surface brightness extended radio structures) as well as efficient, automated RFI flagging, especially at low frequencies. The improvements in telescope sensitivity and data reduction methods played a significant role in motivating us to acquire the new data presented in this paper.

These new data provide a consistent view of the diffuse radio emission on scales of order 10~kpc, with good agreement across a wide frequency range, and reanalysis of the archival data with modern tools shows that the older data are consistent with the new data. We can, therefore, be confident of the radio morphology presented here. However, our new images reveal that the diffuse radio emission is poorly correlated with the hot IGrM, overlapping the cavities but extending beyond them. The emission in and around the cavities has some of the steepest spectral indices, but the spectral index distribution is also poorly correlated with the X-ray morphology. The estimated radiative spectral age for the radio emission (77$\pm$10~Myr) is significantly higher than the timescales previously determined for the formation of the $\sim$5~kpc-scale cavities \citep[$\sim$10-20~Myr,][]{David2009,OSullivanetal11}. However, the latter were estimated assuming cavity expansion at the sound speed and, given the lack of shock heated rims around the cavities, they must be underestimates. The buoyant and sonic timescales for these cavities are comparable, but the refill timescale is significantly longer ($\sim$30-35~Myr) and gas motions may have affected the expansion and rise of the cavities, rendering such dynamical estimates inaccurate. It should also be noted that these dynamical estimates assume the cavities to be located in the plane of the sky, and are thus effectively lower limits; if the cavities are inclined toward the line of sight (i.e., in front of or behind the galaxy core) their dynamical ages would be greater. For the limiting plane-of-sky case, to bring the radio age estimate into agreement with the cavity refill timescale would require a magnetic field of $B\simeq7~\mu G$ or $0.2~\mu G$, which is plausible for a dominant group galaxy. Agreement with the shorter buoyant and sonic timescales would require $B\simeq12-20~\mu G$.

While the inflation of cavities by expanding radio lobes means that we expect radio emission to be confined within the X-ray structure, NGC~5044 is not alone in showing morphological disagreements. Among galaxy groups, NGC~1407 and NGC~3411 (also known as NGC~3402 or SS2b~153) both host large ($\sim$80~kpc) diffuse, steep spectrum ($\alpha=-1.8$ and $-1.4$, respectively) radio structures which surround the group-dominant galaxy and have no corresponding cavity or rim structures in the X-ray \citep{OSullivanetal07,Giacintucci2011,Giacintucci2012}. For NGC~1407, there is evidence that the emission arises from old lobe plasma, and NGC~3411 shows at least some bipolar symmetry that might indicate a similar origin. However, \citet{Giacintucci2011} compared their radio morphology to that of radio mini-halos seen in galaxy clusters. NGC~777 provides a smaller example, with symmetrical $\sim$10~kpc-scale diffuse emission surrounding its radio core, located within an X-ray channel that probably indicates that the radio plasma is the remnant of old, faded lobes \citep{OSullivan2024}. By contrast, the group-dominant galaxies NGC~677 and NGC~1587 \citep{Giacintucci2011,Kolokythasetal18,Kolokythasetal19} host distinctly asymmetric diffuse emission structures on scales of 5-10~kpc, again with no clear correlation with the hot IGrM. These structures are detected at a single frequency and they do not appear consistent with typical jets or lobes. Further observations are needed before their origin can be determined. IC~1860 provides an example of a slightly different phenomenon; it hosts a  $\sim$35~kpc diffuse radio trail extending southeast from its core, but while this is not correlated with any cavities, it is aligned with the innermost part of a sloshing spiral, and probably represents old lobe material drawn out by the IGrM motions \citep{Gastaldello2013}.

Examples of complex radio/X-ray morphologies in galaxy clusters include 2A~0335+096, which contains a complex collection of at least five cavities  whose development may have been affected by sloshing motions in the ICM \citep{Sanders2009}. An inner double radio source is closely correlated with two of these, but diffuse low frequency radio emission overlaps the remaining three and extends well beyond \citep{Birzanetal20,Kokotanekov2017}. Perhaps most famously, ultra-deep high resolution X-ray mapping of the core of the Perseus cluster has revealed multiple cavities whose positions and morphologies have likely been affected by sloshing motions in the ICM. Two pairs of cavities are clearly correlated with the inner and outer radio lobes of NGC~1275, but diffuse radio emission extends well beyond those structures, with filaments linking them to the older "ghost" cavities, and eventually merging with the surrounding mini halo \citep{vanWeeren2024,Kokotanekov2017}. 

These examples make clear that extended radio structures in group and cluster-central galaxies are often more complex than simple cavity-filling lobes. The origin of the more extended emission is not always clear. In cool core clusters, the AGN is not always the dominant source of extended radio emission. Centrally located steep spectrum diffuse radio sources, known as mini-halos, are observed to extend up to 500 kpc, and are often bounded by sloshing fronts in the ICM. The exact nature of the processes responsible for the relativistic electron population in these sources is still under debate \citep[e.g.,][]{Giacintucci2019,Ignesti2020}, but ICM turbulence associated with sloshing motions appears to play a role. However, in groups the gas turbulent motions are thought to be insufficient \citep{Brunetti2014}. No mini-halos have yet been observed in groups, and the emission we observe in NGC~5044 is not consistent with a mini-halo;  its size, radio power, and radio emissivity are not compatible with the known relationship for mini-halos \citep{Murgia2009}, and the radio and X-ray surface brightness are not correlated. 

Perhaps a more likely possibility is that the radio emission reveals plasma, originally in radio jets or lobes, but which has now diffused out into the surrounding IGrM. The deep \chandra\ observations of NGC~5044 have revealed its core to be highly structured, with cavities, channels and filaments of thermal plasma probably shaped by the IGrM ``weather'', driven by the combination of AGN heating, buoyant forces, and sloshing \citep[e.g.,][]{David2011,David2017}. The overall correlation between the H$\alpha$ nebula and the diffuse radio structure (Figure\,\ref{Xray_residual} bottom) could also indicate that the same gas motions shape the two. There is extensive observational evidence of the impact of sloshing on the development of central radio sources.  In addition to the examples given above, NGC~507 provides a prime example of a galaxy group where sloshing has affected both lobe shape and the morphology of plasma filaments which appear to trace material leaking from those lobes \citep{Brienza2022}. Disruption of radio structures by ICM/IGrM motions is also predicted by some simulations \citep[e.g.,][]{Morsony2010,BourneSijacki2017}, with weaker jet sources (such as that in NGC~5044) expected to be more strongly affected \citep{BourneSijacki2021}. However, the processes involved are not sufficiently well understood for us to know whether ``weather" in the core of NGC~5044 could drive the degree of mixing necessary to produce the radio morphology, and the poor correlation with the X-ray morphology, which we observe.

\subsection{Cold gas content: comparison with other systems}
Although a significant fraction of early-type galaxies are found to contain cold gas reservoirs, the mass of cold gas is generally lower than that found in late-type galaxies of comparable mass, and gas content is strongly correlated with the environment, with satellite galaxies in dense environments typically hosting smaller quantities of more disturbed gas \citep{Young2011,Serra2012,Davis2019}. Environmental effects have been found on the group scale, with the cold gas content of individual galaxies and of the group as a whole declining with increasing mass \citep{Kilbornetal09,Desjardinsetal14,Odekonetal16}. There are relatively few systematic comparisons of atomic and molecular gas content, but \citet{Welch2010} found that M$_{H_2}$/M$_{H\textsc{i}}$ in early-type field galaxies tend to cluster around either 0.05 or 0.5, and \citet{ObreschkowRawlings2009} report M$_{H_2}$/M$_{H\textsc{i}}$$\sim$0.5 for ellipticals. The lower limit we derive for NGC~5044, M$_{H_2}$/M$_{H\textsc{i}}$$>$1.76, is consistent with a handful of early-type galaxies at the high end of the range found by these previous studies.

There is morphological and kinematic evidence that many early-type galaxies acquired their cool gas through mergers with, or tidal stripping of,  gas-rich neighbors \citep[e.g.,][]{Serra2012, Maccagni2023}. Some group-dominant galaxies also have the disturbed gas structures characteristic of such interactions \citep[e.g.,][]{OSullivanetal18,Olivares2022}, but the filamentary nebula in NGC~5044 shows very clearly that its cold gas is the product of cooling from the IGrM. \citet{Rose2024} point out that in group and cluster-dominant galaxies, molecular absorption lines seem to arise from two populations of molecular gas clouds. Clouds with low velocity dispersion, small velocity offsets from the galaxy, and a lack of any systemic velocity flow are likely located in stable rotating disks. Clouds with higher velocity dispersions tend to show higher velocity offsets, and in the \citet{Rose2024} 
sample are universally redshifted, indicating that they are infalling material located closer to the AGN.

Of the other group-dominant galaxies in which \Hi\ absorption has been detected, hydrogen column densities in the range $N_{\rm H\textsc{i},abs}$=$2.5-7\times$10$^{18}$~T$_{\rm spin}$~cm$^{-2}$ are found in NGC~4261 and NGC~315 \citep{JaffeMcNamara94,Morgantietal09}, comparable with if somewhat greater than the column density we observe in NGC~5044. NGC~4261 and NGC~6868 both show line profiles with components centered around the systemic velocity \citep{Rose2024,JaffeMcNamara94}. Both also show rotating disks in molecular emission, suggesting that the \Hi\ is likely associated with the disk  \citep[e.g.,][]{Ruffa2019, Ruffa2019a}. NGC~315 also hosts a molecular gas disk \citep{Boizelle2021} but has a more complex \Hi\ absorption structure, with a narrow (FWHM=2.5\kmps) absorption line redshifted by $\sim$490\kmps\ relative to the systemic velocity \citep{Dressel1983} and a somewhat broader (FWHM=80\kmps) absorption component redshifted by only $\sim$80\kmps\ \citep{Morgantietal09}. \Hi\ is also seen in emission a few kiloparsecs from the NGC~315 nucleus, at velocities extending up to that of the narrow absorption line. \cite{Morgantietal09} conclude that the broader absorption likely arises from gas close to the nucleus, and perhaps in the nuclear disk, while the narrow line probably traces an infalling gas cloud located further away, either originating from IGrM cooling, or acquired via tidal stripping. 

NGC~5044 is different from these systems. The redshifted velocity and moderately broadened velocity dispersion of the CO absorption line has been interpreted as indicating that the molecular clouds responsible are not part of a disk, and are likely infalling toward the AGN \citep{David2014,Schellenberger2020,Rose2024}. Based on the velocity of the clouds and their evaporation timescale, \citet{Schellenberger2021} argued that they are very likely to be within the gravitational sphere of influence of the SMBH, which they estimated as having a radius $\sim$22~pc, based on a black hole mass \Msmbh = 2.3$\times$10$^8$\Msol. \cite{Diniz2017} find a black mole mass an order of magnitude higher, \Msmbh = 1.8$\times$10$^9$\Msol, and \citet{Schellenberger2024} find comparable values based on modelling of the radio to sub-millimeter spectral energy distribution. This strengthens the original argument, making it near-certain that the CO and \hi absorption features arise from gas clouds well inside the SMBH sphere of influence.

The molecular-to-atomic gas mass ratio we find for NGC~5044 would be unusually high if it had acquired its gas from another galaxy, but is consistent with the prediction of the chaotic cold accretion (CCA) model advanced by \cite{Gaspari2017}. From simulations of the thermally unstable IGrM in the central 2~kpc of a group-dominant galaxy (using a region and parameters chosen to allow comparison with NGC~5044), \cite{Gaspari2017} found a mass ratio M$_{H_2}$/M$_{H\textsc{i}}$ increasing from $\sim$1 at early times to $\sim$4 after 70-80~Myr of cooling. The mass of the warm ionized component was a factor $\sim$10 lower than that of the neutral atomic gas. Interestingly, this suggests that it should be practical to detect the neutral phase in emission in NGC~5044. If \Hi\ is present in the quantities the CCA model predicts, we would only need to improve on the mass sensitivity of our MeerKAT observation by a factor $\sim$2.3 to detect it.


\section{Conclusions}
We reported wideband multifrequency radio observations (continuum and \hi) of NGC~5044, the dominant elliptical of the X-ray brightest group in the sky. These observations were conducted with the uGMRT (300-850~MHz) and MeerKAT (850~MHz to 1.67~GHz). Our new observations reveal diffuse radio emission extending to $\sim$25~kpc from the radio core. This emission is only weakly correlated with the IGrM morphology and extends well beyond the known X-ray cavities. It shows a strongly curved radio spectrum, relatively flat at low frequencies ($\rm \alpha_{380\,\rm MHz}^{986\,\rm MHz}=-0.70\pm0.05$) but steepening at high frequencies ($\alpha_{\rm 986\,MHz}^{1.56\,\rm GHz}=-1.53\pm0.06$). Using a revised equipartition magnetic field of $B=1.4~\mu G$, the estimated radiative spectral age is  around $77-80$~Myr. We suggest that this newly detected diffuse radio emission likely originated as relativistic plasma of the AGN jets or lobes, which has since diffused into the IGrM by gas motions or ``weather'' driven by the combination of AGN heating, tidal sloshing, and radiative cooling. NGC~5044 is not alone among groups or clusters in hosting complex diffuse radio emission which is not confined within clear cavities. As the sensitivity of radio observations improves, it is likely that such emission will be discovered in an increasing number of systems.

We also report the first detection (at 3.8 $\sigma$ level) of neutral atomic gas in the cooling region of NGC~5044, in the form of a double-structured \hi line seen in absorption against the radio core of NGC~5044. The velocities of the \hi peaks are consistent with those of the CO(2-1) absorption lines previously reported in the galaxy, suggesting that both sets of lines trace different density zones of the same gas clouds. The \Hi\ lines show moderately broader velocity dispersions compared to the CO(2-1), consistent with our expectations in this scenario. We find the column density and velocity width of the \Hi\ lines to be comparable to those seen in the handful of other cool-core systems where \Hi\ is detected in absorption, and conclude that the clouds are likely inside the sphere of influence of the central SMBH and infalling onto the AGN, as suggested by previous studies. From the non-detection of the \hi emission in the core of NGC~5044, we place an upper limit on the atomic mass of $M_{\hi}< 5.4 \times 10^{7} \, M_{\odot}$ in the central 2.2~kpc (15\arcs). The molecular-to-atomic mass ratio ($\rm M_{H_2}/M_{H\textsc{i}}\gtsim1.7$) aligns with predictions of the chaotic cold accretion model, which implies that only modest increases in sensitivity may be needed to detect the \Hi\ emission from the neutral atomic phase of the cooling flow in NGC~5044. We also obtained an upper for the spin temperature of $\rm T_{\rm spin} \leq 950 \, \rm K$, suggesting that the observed \hi is in the cold neutral phase. 

\section*{acknowledgments}
KR acknowledges the Smithsonian Combined Support for Life on a Sustainable Planet, Science, and Research administered by the Office of the Under Secretary for Science and Research. TD acknowledges funding from an NWO Rubicon Fellowship, project number 019.231EN.001. WF acknowledges support from the Smithsonian Institution, the Chandra High Resolution Camera Project through NASA contract NAS8-0306, NASA Grant 80NSSC19K0116, and Chandra Grant GO1-22132X. Basic research in radio astronomy at the U.S. Naval Research Laboratory is supported by 6.1 Base Funding. We acknowledge the use of the {\it ilifu} cloud computing facility - www.ilifu.ac.za, a partnership between the University of Cape Town, the University of the Western Cape, Stellenbosch University, Sol Plaatje University and the Cape Peninsula University of Technology. The ilifu facility is supported by contributions from the Inter-University Institute for Data Intensive Astronomy (IDIA - a partnership between the University of Cape Town, the University of Pretoria and the University of the Western Cape), the Computational Biology division at UCT and the Data Intensive Research Initiative of South Africa (DIRISA). This work made use of the CARTA (Cube Analysis and Rendering Tool for Astronomy) software. The authors thank the staff of the MeerKAT observatory for their help with the observations presented in this work.  The MeerKAT telescope is operated by the South African Radio Astronomy Observatory, which is a facility of the National Research Foundation, an agency of the Department of Science and Innovation. We thank the staff of the GMRT that made these observations possible. GMRT is run by the National Centre for Radio Astrophysics of the Tata Institute of Fundamental Research.

\facilities{MeerKAT}, {uGMRT}, {CXO}

\software{CARACal \citep{caracal2020}, AOflagger \citep{Offringa2010}, WSClean \citep{Offringa2014}, SPAM \citep{Intema2009}, CARTA \citep{CARTA2021}, DS9 \citep{ds9}, SoFiA \citep{Serra2015}, SoFiA Image Pipeline, Astropy \citep{astropy2013, astropy2018}, APLpy \citep{aplpy}, Matplotlib \citep{matplotlib}}
\appendix
\restartappendixnumbering 
\renewcommand{\thefigure}{A.\arabic{figure}}

\begin{figure*}[ht]
\centering
\includegraphics[width=1.0\textwidth]{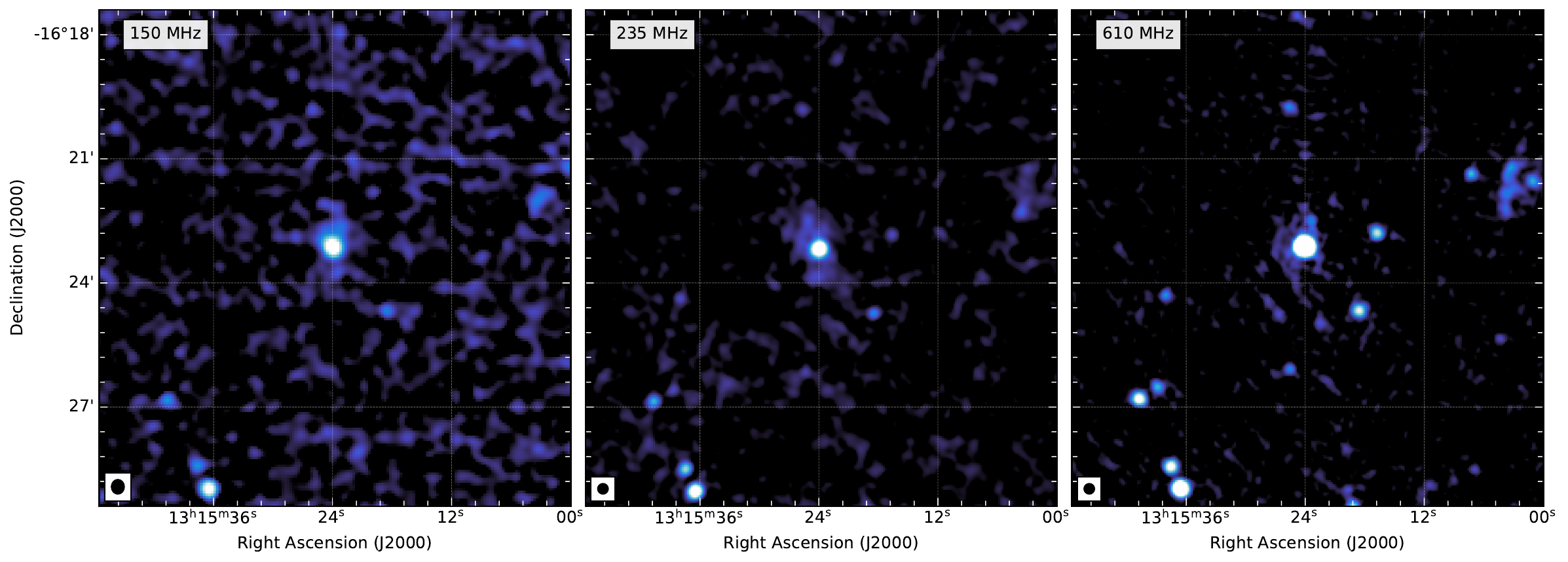}
\caption{GMRT archival 150~MHz (20\arcsec~resolution), 235 MHz (15\arcsec~resolution) and 610 MHz (15\arcsec~resolution) images. The noise level is $1\,\rm mJy\,beam^{-1}$, $228\,\mu \rm Jy\,beam^{-1}$, $ 47\,\mu \rm Jy\,beam^{-1}$ at 150 MHz, 235 MHz, and 610 MHz, respectively. The beam size is indicated in the bottom left corner of the image. }
\label{fig:appendix1}
\end{figure*}

\begin{figure*}[ht]
\centering
\includegraphics[width=0.9\textwidth]{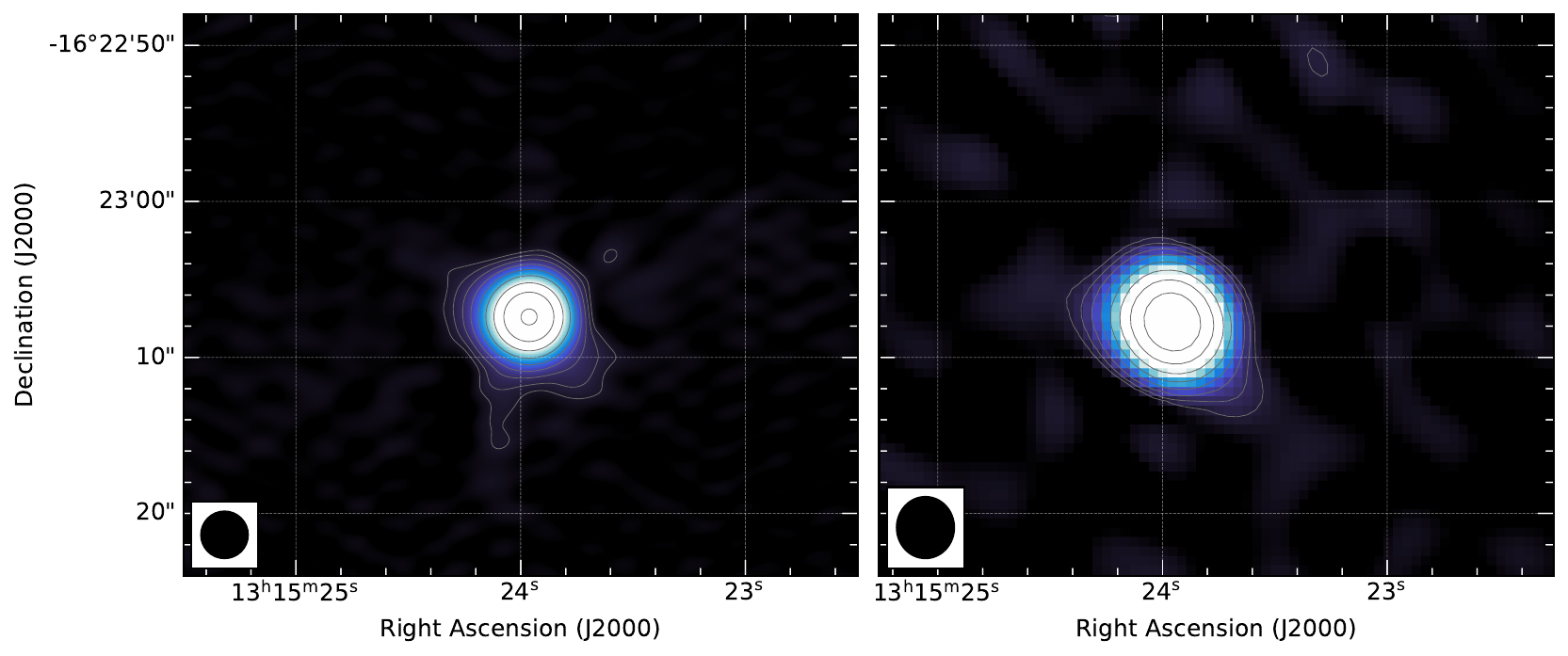}
\caption{High resolution VLA AnB configuration ($3\arcsec$) and  MeerKAT 1.56~GHz ($3.9\arcsec \times 3.7\arcsec$) images. The noise levels are $20\,\rm \mu Jy\,beam^{-1}$ and $ 23\,\mu \rm Jy\,beam^{-1}$ at 1.5~GHz and 1.28 GHz, respectively. The images are created using Briggs weighting with $\rm {\tt roboust}=-2$ and an inner uv cut. Contours are drawn at $[1, 2, 4, 8 ...]\times 5.0\sigma_{\rm rms}$. The beam size is indicated in the bottom left corner of each image.}
\label{fig:appendix2}
\end{figure*}

In Figure\,\ref{fig:appendix1}, we present images of NGC\,5044 obtained from archival GMRT data at 150~MHz (proposal code: 22\_008), 235~MHz (proposal code: 17\_034), and 610~MHz (proposal code: 17\_034). Data reduction was carried out in {\tt SPAM}. The overall morphology of the radio emission around NGC 5044 is consistent with that observed in our new deep wideband uGMRT and MeerKAT images. We emphasize that the noise levels in our archival GMRT images are comparable to those in previously published images.  In Figure\,\ref{fig:appendix2}, we present the highest-resolution MeerKAT L-band and VLA AnB configuration images (proposal code: 15A-243). The image is obtained by combining four BnA configuration data sets.  

\bibliographystyle{aasjournal}
\bibliography{ref}

\begin{thebibliography}{}
\expandafter\ifx\csname natexlab\endcsname\relax\def\natexlab#1{#1}\fi
\providecommand{\url}[1]{\href{#1}{#1}}
\providecommand{\dodoi}[1]{doi:~\href{http://doi.org/#1}{\nolinkurl{#1}}}
\providecommand{\doeprint}[1]{\href{http://ascl.net/#1}{\nolinkurl{http://ascl.net/#1}}}
\providecommand{\doarXiv}[1]{\href{https://arxiv.org/abs/#1}{\nolinkurl{https://arxiv.org/abs/#1}}}

\bibitem[{Allison(2021)}]{Allison2021}
Allison, J.~R. 2021, Monthly Notices of the Royal Astronomical Society, 503,
  985, \dodoi{10.1093/mnras/stab518}

\bibitem[{{Astropy Collaboration} {et~al.}(2013){Astropy Collaboration},
  {Robitaille}, {Tollerud}, {Greenfield}, {Droettboom}, {Bray}, {Aldcroft},
  {Davis}, {Ginsburg}, {Price-Whelan}, {Kerzendorf}, {Conley}, {Crighton},
  {Barbary}, {Muna}, {Ferguson}, {Grollier}, {Parikh}, {Nair}, {Unther},
  {Deil}, {Woillez}, {Conseil}, {Kramer}, {Turner}, {Singer}, {Fox}, {Weaver},
  {Zabalza}, {Edwards}, {Azalee Bostroem}, {Burke}, {Casey}, {Crawford},
  {Dencheva}, {Ely}, {Jenness}, {Labrie}, {Lim}, {Pierfederici}, {Pontzen},
  {Ptak}, {Refsdal}, {Servillat}, \& {Streicher}}]{astropy2013}
{Astropy Collaboration}, {Robitaille}, T.~P., {Tollerud}, E.~J., {et~al.} 2013,
  \aap, 558, A33, \dodoi{10.1051/0004-6361/201322068}

\bibitem[{{Astropy Collaboration} {et~al.}(2018){Astropy Collaboration},
  {Price-Whelan}, {Sip{\H{o}}cz}, {G{\"u}nther}, {Lim}, {Crawford}, {Conseil},
  {Shupe}, {Craig}, {Dencheva}, {Ginsburg}, {VanderPlas}, {Bradley},
  {P{\'e}rez-Su{\'a}rez}, {de Val-Borro}, {Aldcroft}, {Cruz}, {Robitaille},
  {Tollerud}, {Ardelean}, {Babej}, {Bach}, {Bachetti}, {Bakanov}, {Bamford},
  {Barentsen}, {Barmby}, {Baumbach}, {Berry}, {Biscani}, {Boquien}, {Bostroem},
  {Bouma}, {Brammer}, {Bray}, {Breytenbach}, {Buddelmeijer}, {Burke},
  {Calderone}, {Cano Rodr{\'\i}guez}, {Cara}, {Cardoso}, {Cheedella}, {Copin},
  {Corrales}, {Crichton}, {D'Avella}, {Deil}, {Depagne}, {Dietrich}, {Donath},
  {Droettboom}, {Earl}, {Erben}, {Fabbro}, {Ferreira}, {Finethy}, {Fox},
  {Garrison}, {Gibbons}, {Goldstein}, {Gommers}, {Greco}, {Greenfield},
  {Groener}, {Grollier}, {Hagen}, {Hirst}, {Homeier}, {Horton}, {Hosseinzadeh},
  {Hu}, {Hunkeler}, {Ivezi{\'c}}, {Jain}, {Jenness}, {Kanarek}, {Kendrew},
  {Kern}, {Kerzendorf}, {Khvalko}, {King}, {Kirkby}, {Kulkarni}, {Kumar},
  {Lee}, {Lenz}, {Littlefair}, {Ma}, {Macleod}, {Mastropietro}, {McCully},
  {Montagnac}, {Morris}, {Mueller}, {Mumford}, {Muna}, {Murphy}, {Nelson},
  {Nguyen}, {Ninan}, {N{\"o}the}, {Ogaz}, {Oh}, {Parejko}, {Parley}, {Pascual},
  {Patil}, {Patil}, {Plunkett}, {Prochaska}, {Rastogi}, {Reddy Janga},
  {Sabater}, {Sakurikar}, {Seifert}, {Sherbert}, {Sherwood-Taylor}, {Shih},
  {Sick}, {Silbiger}, {Singanamalla}, {Singer}, {Sladen}, {Sooley},
  {Sornarajah}, {Streicher}, {Teuben}, {Thomas}, {Tremblay}, {Turner},
  {Terr{\'o}n}, {van Kerkwijk}, {de la Vega}, {Watkins}, {Weaver}, {Whitmore},
  {Woillez}, {Zabalza}, \& {Astropy Contributors}}]{astropy2018}
{Astropy Collaboration}, {Price-Whelan}, A.~M., {Sip{\H{o}}cz}, B.~M., {et~al.}
  2018, \aj, 156, 123, \dodoi{10.3847/1538-3881/aabc4f}

\bibitem[{{B{\^i}rzan} {et~al.}(2008){B{\^i}rzan}, {McNamara}, {Nulsen},
  {Carilli}, \& {Wise}}]{Birzanetal08}
{B{\^i}rzan}, L., {McNamara}, B.~R., {Nulsen}, P.~E.~J., {Carilli}, C.~L., \&
  {Wise}, M.~W. 2008, \apj, 686, 859, \dodoi{10.1086/591416}

\bibitem[{{B{\^\i}rzan} {et~al.}(2020){B{\^\i}rzan}, {Rafferty}, {Br{\"u}ggen},
  {Botteon}, {Brunetti}, {Cuciti}, {Edge}, {Morganti}, {R{\"o}ttgering}, \&
  {Shimwell}}]{Birzanetal20}
{B{\^\i}rzan}, L., {Rafferty}, D.~A., {Br{\"u}ggen}, M., {et~al.} 2020, \mnras,
  496, 2613, \dodoi{10.1093/mnras/staa1594}

\bibitem[{{Boizelle} {et~al.}(2021){Boizelle}, {Walsh}, {Barth}, {Buote},
  {Baker}, {Darling}, {Ho}, {Cohn}, \& {Kabasares}}]{Boizelle2021}
{Boizelle}, B.~D., {Walsh}, J.~L., {Barth}, A.~J., {et~al.} 2021, \apj, 908,
  19, \dodoi{10.3847/1538-4357/abd24d}

\bibitem[{{Bourne} \& {Sijacki}(2017)}]{BourneSijacki2017}
{Bourne}, M.~A., \& {Sijacki}, D. 2017, \mnras, 472, 4707,
  \dodoi{10.1093/mnras/stx2269}

\bibitem[{{Bourne} \& {Sijacki}(2021)}]{BourneSijacki2021}
---. 2021, \mnras, 506, 488, \dodoi{10.1093/mnras/stab1662}

\bibitem[{{Brienza} {et~al.}(2022){Brienza}, {Lovisari}, {Rajpurohit},
  {Bonafede}, {Gastaldello}, {Murgia}, {Vazza}, {Bonnassieux}, {Botteon},
  {Brunetti}, {Drabent}, {Hardcastle}, {Pasini}, {Riseley}, {R{\"o}ttgering},
  {Shimwell}, {Simionescu}, \& {van Weeren}}]{Brienza2022}
{Brienza}, M., {Lovisari}, L., {Rajpurohit}, K., {et~al.} 2022, \aap, 661, A92,
  \dodoi{10.1051/0004-6361/202142579}

\bibitem[{{Briggs}(1995)}]{Briggs1995}
{Briggs}, D.~S. 1995, PhD thesis, New Mexico Institute of Mining and Technology

\bibitem[{{Brunetti} \& {Jones}(2014)}]{Brunetti2014}
{Brunetti}, G., \& {Jones}, T.~W. 2014, International Journal of Modern Physics
  D, 23, 1430007, \dodoi{10.1142/S0218271814300079}

\bibitem[{{Brunetti} {et~al.}(1997){Brunetti}, {Setti}, \&
  {Comastri}}]{Brunetti1997}
{Brunetti}, G., {Setti}, G., \& {Comastri}, A. 1997, \aap, 325, 898,
  \dodoi{10.48550/arXiv.astro-ph/9704162}

\bibitem[{{Buote} {et~al.}(2003){Buote}, {Lewis}, {Brighenti}, \&
  {Mathews}}]{Buote2003a}
{Buote}, D.~A., {Lewis}, A.~D., {Brighenti}, F., \& {Mathews}, W.~G. 2003,
  \apj, 594, 741, \dodoi{10.1086/377094}

\bibitem[{{Churazov} {et~al.}(2001){Churazov}, {Br{\"u}ggen}, {Kaiser},
  {B{\"o}hringer}, \& {Forman}}]{Churazov2001}
{Churazov}, E., {Br{\"u}ggen}, M., {Kaiser}, C.~R., {B{\"o}hringer}, H., \&
  {Forman}, W. 2001, \apj, 554, 261, \dodoi{10.1086/321357}

\bibitem[{{Churazov} {et~al.}(2000){Churazov}, {Forman}, {Jones}, \&
  {B{\"o}hringer}}]{Churazov2000}
{Churazov}, E., {Forman}, W., {Jones}, C., \& {B{\"o}hringer}, H. 2000, \aap,
  356, 788, \dodoi{10.48550/arXiv.astro-ph/0002375}

\bibitem[{{Churazov} {et~al.}(2002){Churazov}, {Sunyaev}, {Forman}, \&
  {B{\"o}hringer}}]{Churazov2002}
{Churazov}, E., {Sunyaev}, R., {Forman}, W., \& {B{\"o}hringer}, H. 2002,
  \mnras, 332, 729, \dodoi{10.1046/j.1365-8711.2002.05332.x}

\bibitem[{Comrie {et~al.}(2021)Comrie, Wang, Hsu, Moraghan, Harris, Pang,
  Pińska, Chiang, Chang, Hwang, Jan, Lin, \& Simmonds}]{CARTA2021}
Comrie, A., Wang, K.-S., Hsu, S.-C., {et~al.} 2021, {CARTA: The Cube Analysis
  and Rendering Tool for Astronomy}, 2.0.0,  Zenodo,
  \dodoi{10.5281/zenodo.4905459}

\bibitem[{{David} {et~al.}(1994){David}, {Jones}, {Forman}, \&
  {Daines}}]{David1994}
{David}, L.~P., {Jones}, C., {Forman}, W., \& {Daines}, S. 1994, \apj, 428,
  544, \dodoi{10.1086/174264}

\bibitem[{{David} {et~al.}(2009){David}, {Jones}, {Forman}, {Nulsen},
  {Vrtilek}, {O'Sullivan}, {Giacintucci}, \& {Raychaudhury}}]{David2009}
{David}, L.~P., {Jones}, C., {Forman}, W., {et~al.} 2009, \apj, 705, 624,
  \dodoi{10.1088/0004-637X/705/1/624}

\bibitem[{{David} {et~al.}(2017){David}, {Vrtilek}, {O'Sullivan}, {Jones},
  {Forman}, \& {Sun}}]{David2017}
{David}, L.~P., {Vrtilek}, J., {O'Sullivan}, E., {et~al.} 2017, \apj, 842, 84,
  \dodoi{10.3847/1538-4357/aa756c}

\bibitem[{{David} {et~al.}(2011){David}, {O'Sullivan}, {Jones}, {Giacintucci},
  {Vrtilek}, {Raychaudhury}, {Nulsen}, {Forman}, {Sun}, \&
  {Donahue}}]{David2011}
{David}, L.~P., {O'Sullivan}, E., {Jones}, C., {et~al.} 2011, \apj, 728, 162,
  \dodoi{10.1088/0004-637X/728/2/162}

\bibitem[{{David} {et~al.}(2014){David}, {Lim}, {Forman}, {Vrtilek}, {Combes},
  {Salome}, {Edge}, {Hamer}, {Jones}, {Sun}, {O'Sullivan}, {Gastaldello},
  {Bardelli}, {Temi}, {Schmitt}, {Ohyama}, {Mathews}, {Brighenti},
  {Giacintucci}, \& {Trung}}]{David2014}
{David}, L.~P., {Lim}, J., {Forman}, W., {et~al.} 2014, \apj, 792, 94,
  \dodoi{10.1088/0004-637X/792/2/94}

\bibitem[{{Davis} {et~al.}(2019){Davis}, {Greene}, {Ma}, {Blakeslee}, {Dawson},
  {Pandya}, {Veale}, \& {Zabel}}]{Davis2019}
{Davis}, T.~A., {Greene}, J.~E., {Ma}, C.-P., {et~al.} 2019, \mnras, 486, 1404,
  \dodoi{10.1093/mnras/stz871}

\bibitem[{Deb {et~al.}(2022)Deb, Verheijen, Poggianti, Moretti, van~der Hulst,
  Vulcani, Ramatsoku, Serra, Healy, Gullieuszik, Bacchini, Ignesti, Müller,
  Zabel, Luber, Jaffë, \& Gitti}]{Tirna_2022}
Deb, T., Verheijen, M. A.~W., Poggianti, B.~M., {et~al.} 2022, Monthly Notices
  of the Royal Astronomical Society, 516, 2683–2696,
  \dodoi{10.1093/mnras/stac2441}

\bibitem[{{Desjardins} {et~al.}(2014){Desjardins}, {Gallagher},
  {Hornschemeier}, {Mulchaey}, {Walker}, {Brandt}, {Charlton}, {Johnson}, \&
  {Tzanavaris}}]{Desjardinsetal14}
{Desjardins}, T.~D., {Gallagher}, S.~C., {Hornschemeier}, A.~E., {et~al.} 2014,
  \apj, 790, 132

\bibitem[{{Diniz} {et~al.}(2017){Diniz}, {Pastoriza}, {Hernandez-Jimenez},
  {Riffel}, {Ricci}, {Steiner}, \& {Riffel}}]{Diniz2017}
{Diniz}, S. I.~F., {Pastoriza}, M.~G., {Hernandez-Jimenez}, J.~A., {et~al.}
  2017, \mnras, 470, 1703, \dodoi{10.1093/mnras/stx1322}

\bibitem[{{Donahue} \& {Voit}(2022)}]{DonahueVoit2022}
{Donahue}, M., \& {Voit}, G.~M. 2022, \physrep, 973, 1,
  \dodoi{10.1016/j.physrep.2022.04.005}

\bibitem[{{Dong} {et~al.}(2010){Dong}, {Rasmussen}, \& {Mulchaey}}]{Dongetal10}
{Dong}, R., {Rasmussen}, J., \& {Mulchaey}, J.~S. 2010, \apj, 712, 883,
  \dodoi{10.1088/0004-637X/712/2/883}

\bibitem[{{Dressel} {et~al.}(1983){Dressel}, {Bania}, \& {Davis}}]{Dressel1983}
{Dressel}, L.~L., {Bania}, T.~M., \& {Davis}, M.~M. 1983, \apjl, 266, L97,
  \dodoi{10.1086/183986}

\bibitem[{{Eckert} {et~al.}(2021){Eckert}, {Gaspari}, {Gastaldello}, {Le Brun},
  \& {O'Sullivan}}]{Eckert2021}
{Eckert}, D., {Gaspari}, M., {Gastaldello}, F., {Le Brun}, A. M.~C., \&
  {O'Sullivan}, E. 2021, Universe, 7, 142, \dodoi{10.3390/universe7050142}

\bibitem[{{Eckert} {et~al.}(2016){Eckert}, {Ettori}, {Coupon}, {Gastaldello},
  {Pierre}, {Melin}, {Le Brun}, {McCarthy}, {Adami}, {Chiappetti}, {Faccioli},
  {Giles}, {Lavoie}, {Lef{\`e}vre}, {Lieu}, {Mantz}, {Maughan}, {McGee},
  {Pacaud}, {Paltani}, {Sadibekova}, {Smith}, \& {Ziparo}}]{Eckert2016}
{Eckert}, D., {Ettori}, S., {Coupon}, J., {et~al.} 2016, \aap, 592, A12

\bibitem[{{Edge}(2001)}]{Edge2001}
{Edge}, A.~C. 2001, \mnras, 328, 762, \dodoi{10.1046/j.1365-8711.2001.04802.x}

\bibitem[{{Fabian}(2012)}]{Fabian12}
{Fabian}, A.~C. 2012, \araa, 50, 455,
  \dodoi{10.1146/annurev-astro-081811-125521}

\bibitem[{{Gaspari} {et~al.}(2017){Gaspari}, {Temi}, \&
  {Brighenti}}]{Gaspari2017}
{Gaspari}, M., {Temi}, P., \& {Brighenti}, F. 2017, \mnras, 466, 677,
  \dodoi{10.1093/mnras/stw3108}

\bibitem[{{Gaspari} {et~al.}(2020){Gaspari}, {Tombesi}, \&
  {Cappi}}]{Gaspari2020}
{Gaspari}, M., {Tombesi}, F., \& {Cappi}, M. 2020, Nature Astronomy, 4, 10,
  \dodoi{10.1038/s41550-019-0970-1}

\bibitem[{{Gastaldello} {et~al.}(2007){Gastaldello}, {Buote}, {Humphrey},
  {Zappacosta}, {Bullock}, {Brighenti}, \& {Mathews}}]{Gastaldello2007}
{Gastaldello}, F., {Buote}, D.~A., {Humphrey}, P.~J., {et~al.} 2007, \apj, 669,
  158, \dodoi{10.1086/521519}

\bibitem[{{Gastaldello} {et~al.}(2009){Gastaldello}, {Buote}, {Temi},
  {Brighenti}, {Mathews}, \& {Ettori}}]{Gastaldello2009}
{Gastaldello}, F., {Buote}, D.~A., {Temi}, P., {et~al.} 2009, \apj, 693, 43,
  \dodoi{10.1088/0004-637X/693/1/43}

\bibitem[{{Gastaldello} {et~al.}(2013){Gastaldello}, {Di Gesu}, {Ghizzardi},
  {Giacintucci}, {Girardi}, {Roediger}, {Rossetti}, {Brighenti}, {Buote},
  {Eckert}, {Ettori}, {Humphrey}, \& {Mathews}}]{Gastaldello2013}
{Gastaldello}, F., {Di Gesu}, L., {Ghizzardi}, S., {et~al.} 2013, \apj, 770,
  56, \dodoi{10.1088/0004-637X/770/1/56}

\bibitem[{{Giacintucci} {et~al.}(2019){Giacintucci}, {Markevitch}, {Cassano},
  {Venturi}, {Clarke}, {Kale}, \& {Cuciti}}]{Giacintucci2019}
{Giacintucci}, S., {Markevitch}, M., {Cassano}, R., {et~al.} 2019, \apj, 880,
  70, \dodoi{10.3847/1538-4357/ab29f1}

\bibitem[{{Giacintucci} {et~al.}(2009){Giacintucci}, {Vrtilek}, {O'Sullivan},
  {Raychaudhury}, {David}, {Venturi}, {Athreya}, \& {Gitti}}]{Giacintucci2009}
{Giacintucci}, S., {Vrtilek}, J.~M., {O'Sullivan}, E., {et~al.} 2009, in
  American Institute of Physics Conference Series, Vol. 1201, The Monster's
  Fiery Breath: Feedback in Galaxies, Groups, and Clusters, ed. S.~{Heinz} \&
  E.~{Wilcots} (AIP), 229--232, \dodoi{10.1063/1.3293043}

\bibitem[{{Giacintucci} {et~al.}(2011){Giacintucci}, {O'Sullivan}, {Vrtilek},
  {David}, {Raychaudhury}, {Venturi}, {Athreya}, {Clarke}, {Murgia},
  {Mazzotta}, {Gitti}, {Ponman}, {Ishwara-Chandra}, {Jones}, \&
  {Forman}}]{Giacintucci2011}
{Giacintucci}, S., {O'Sullivan}, E., {Vrtilek}, J., {et~al.} 2011, \apj, 732,
  95, \dodoi{10.1088/0004-637X/732/2/95}

\bibitem[{{Giacintucci} {et~al.}(2012){Giacintucci}, {O'Sullivan}, {Clarke},
  {Murgia}, {Vrtilek}, {Venturi}, {David}, {Raychaudhury}, \&
  {Athreya}}]{Giacintucci2012}
{Giacintucci}, S., {O'Sullivan}, E., {Clarke}, T.~E., {et~al.} 2012, \apj, 755,
  172, \dodoi{10.1088/0004-637X/755/2/172}

\bibitem[{{Gitti} {et~al.}(2012){Gitti}, {Brighenti}, \& {McNamara}}]{Gitti12}
{Gitti}, M., {Brighenti}, F., \& {McNamara}, B.~R. 2012, Advances in Astronomy,
  2012, 950641, \dodoi{10.1155/2012/950641}

\bibitem[{{Grossov{\'a}} {et~al.}(2022){Grossov{\'a}}, {Werner}, {Massaro},
  {Lakhchaura}, {Pl{\v{s}}ek}, {Gab{\'a}nyi}, {Rajpurohit}, {Canning},
  {Nulsen}, {O'Sullivan}, {Allen}, \& {Fabian}}]{Romana2022}
{Grossov{\'a}}, R., {Werner}, N., {Massaro}, F., {et~al.} 2022, \apjs, 258, 30,
  \dodoi{10.3847/1538-4365/ac366c}

\bibitem[{{Healy} {et~al.}(2021){Healy}, {Deb}, {Verheijen}, {Blyth}, {Serra},
  {Ramatsoku}, \& {Vulcani}}]{Tirna2021}
{Healy}, J., {Deb}, T., {Verheijen}, M.~A.~W., {et~al.} 2021, \aap, 654, A173,
  \dodoi{10.1051/0004-6361/202141377}

\bibitem[{{Hogan}(2014)}]{Hogan14}
{Hogan}, M.~T. 2014, PhD thesis, Durham University, UK

\bibitem[{Hunter(2007)}]{matplotlib}
Hunter, J.~D. 2007, Computing in Science \& Engineering, 9, 90,
  \dodoi{10.1109/MCSE.2007.55}

\bibitem[{{Ignesti} {et~al.}(2020){Ignesti}, {Brunetti}, {Gitti}, \&
  {Giacintucci}}]{Ignesti2020}
{Ignesti}, A., {Brunetti}, G., {Gitti}, M., \& {Giacintucci}, S. 2020, \aap,
  640, A37, \dodoi{10.1051/0004-6361/201937207}

\bibitem[{{Intema} {et~al.}(2009){Intema}, {van der Tol}, {Cotton}, {Cohen},
  {van Bemmel}, \& {R{\"o}ttgering}}]{Intema2009}
{Intema}, H.~T., {van der Tol}, S., {Cotton}, W.~D., {et~al.} 2009, \aap, 501,
  1185, \dodoi{10.1051/0004-6361/200811094}

\bibitem[{{Jaffe} \& {McNamara}(1994)}]{JaffeMcNamara94}
{Jaffe}, W., \& {McNamara}, B.~R. 1994, \apj, 434, 110, \dodoi{10.1086/174708}

\bibitem[{{Jaffe} \& {Perola}(1973)}]{Jaffe1973}
{Jaffe}, W.~J., \& {Perola}, G.~C. 1973, \aap, 26, 423

\bibitem[{{Joye} \& {Mandel}(2003)}]{ds9}
{Joye}, W.~A., \& {Mandel}, E. 2003, in Astronomical Society of the Pacific
  Conference Series, Vol. 295, Astronomical Data Analysis Software and Systems
  XII, ed. H.~E. {Payne}, R.~I. {Jedrzejewski}, \& R.~N. {Hook}, 489

\bibitem[{{J{\'o}zsa} {et~al.}(2020){J{\'o}zsa}, {White}, {Thorat}, {Smirnov},
  {Serra}, {Ramatsoku}, {Ramaila}, {Perkins}, {Moln{\'a}r}, {Makhathini},
  {Maccagni}, {Kleiner}, {Kamphuis}, {Hugo}, {de Blok}, \&
  {Andati}}]{caracal2020}
{J{\'o}zsa}, G. I.~G., {White}, S.~V., {Thorat}, K., {et~al.} 2020, {CARACal:
  Containerized Automated Radio Astronomy Calibration pipeline}, Astrophysics
  Source Code Library, record ascl:2006.014.
\newblock \doeprint{2006.014}

\bibitem[{{Kaneda} {et~al.}(2008){Kaneda}, {Onaka}, {Sakon}, {Kitayama},
  {Okada}, \& {Suzuki}}]{Kaneda2008}
{Kaneda}, H., {Onaka}, T., {Sakon}, I., {et~al.} 2008, \apj, 684, 270,
  \dodoi{10.1086/590243}

\bibitem[{{Kenyon} {et~al.}(2018){Kenyon}, {Smirnov}, {Grobler}, \&
  {Perkins}}]{Kenyon2018}
{Kenyon}, J.~S., {Smirnov}, O.~M., {Grobler}, T.~L., \& {Perkins}, S.~J. 2018,
  \mnras, 478, 2399, \dodoi{10.1093/mnras/sty1221}

\bibitem[{{Kilborn} {et~al.}(2009){Kilborn}, {Forbes}, {Barnes}, {Koribalski},
  {Brough}, \& {Kern}}]{Kilbornetal09}
{Kilborn}, V.~A., {Forbes}, D.~A., {Barnes}, D.~G., {et~al.} 2009, \mnras, 400,
  1962

\bibitem[{{Kirkpatrick} {et~al.}(2011){Kirkpatrick}, {McNamara}, \&
  {Cavagnolo}}]{Kirkpatricketal11}
{Kirkpatrick}, C.~C., {McNamara}, B.~R., \& {Cavagnolo}, K.~W. 2011, \apj, 731,
  L23, \dodoi{10.1088/2041-8205/731/2/L23}

\bibitem[{{Kokotanekov} {et~al.}(2017){Kokotanekov}, {Wise}, {Heald}, {McKean},
  {B{\^\i}rzan}, {Rafferty}, {Godfrey}, {de Vries}, {Intema}, {Broderick},
  {Hardcastle}, {Bonafede}, {Clarke}, {van Weeren}, {R{\"o}ttgering}, {Pizzo},
  {Iacobelli}, {Orr{\'u}}, {Shulevski}, {Riseley}, {Breton},
  {Nikiel-Wroczy{\'n}ski}, {Sridhar}, {Stewart}, {Rowlinson}, {van der Horst},
  {Harwood}, {G{\"u}rkan}, {Carbone}, {Pandey-Pommier}, {Tasse}, {Scaife},
  {Pratley}, {Ferrari}, {Croston}, {Pandey}, {Jurusik}, \&
  {Mulcahy}}]{Kokotanekov2017}
{Kokotanekov}, G., {Wise}, M., {Heald}, G.~H., {et~al.} 2017, \aap, 605, A48,
  \dodoi{10.1051/0004-6361/201730940}

\bibitem[{{Kolokythas} {et~al.}(2019){Kolokythas}, {O'Sullivan}, {Intema},
  {Raychaudhury}, {Babul}, {Giacintucci}, \& {Gitti}}]{Kolokythasetal19}
{Kolokythas}, K., {O'Sullivan}, E., {Intema}, H., {et~al.} 2019, \mnras, 489,
  2488, \dodoi{10.1093/mnras/stz2082}

\bibitem[{{Kolokythas} {et~al.}(2018){Kolokythas}, {O'Sullivan},
  {Raychaudhury}, {Giacintucci}, {Gitti}, \& {Babul}}]{Kolokythasetal18}
{Kolokythas}, K., {O'Sullivan}, E., {Raychaudhury}, S., {et~al.} 2018, \mnras,
  481, 1550, \dodoi{10.1093/mnras/sty2030}

\bibitem[{{Lakhchaura} {et~al.}(2018){Lakhchaura}, {Werner}, {Sun}, {Canning},
  {Gaspari}, {Allen}, {Connor}, {Donahue}, \& {Sarazin}}]{Lakhchauraetal18}
{Lakhchaura}, K., {Werner}, N., {Sun}, M., {et~al.} 2018, \mnras, 481, 4472,
  \dodoi{10.1093/mnras/sty2565}

\bibitem[{{Lovisari} {et~al.}(2015){Lovisari}, {Reiprich}, \&
  {Schellenberger}}]{Lovisari2015}
{Lovisari}, L., {Reiprich}, T.~H., \& {Schellenberger}, G. 2015, \aap, 573,
  A118, \dodoi{10.1051/0004-6361/201423954}

\bibitem[{{Maccagni} {et~al.}(2023){Maccagni}, {Ruffa}, {Loni}, {Prandoni},
  {Ragusa}, {Kleiner}, {Serra}, {Iodice}, \& {Spavone}}]{Maccagni2023}
{Maccagni}, F.~M., {Ruffa}, I., {Loni}, A., {et~al.} 2023, \aap, 675, A59,
  \dodoi{10.1051/0004-6361/202346521}

\bibitem[{{McNamara} \& {Nulsen}(2012)}]{McNamaraNulsen12}
{McNamara}, B.~R., \& {Nulsen}, P.~E.~J. 2012, New Journal of Physics, 14,
  055023, \dodoi{10.1088/1367-2630/14/5/055023}

\bibitem[{{Meyer} {et~al.}(2017){Meyer}, {Robotham}, {Obreschkow}, {Westmeier},
  {Duffy}, \& {Staveley-Smith}}]{Meyer2017}
{Meyer}, M., {Robotham}, A., {Obreschkow}, D., {et~al.} 2017, \pasa, 34, 52,
  \dodoi{10.1017/pasa.2017.31}

\bibitem[{{Morganti} \& {Oosterloo}(2018)}]{MorgantiOosterloo18}
{Morganti}, R., \& {Oosterloo}, T. 2018, \aapr, 26, 4,
  \dodoi{10.1007/s00159-018-0109-x}

\bibitem[{{Morganti} {et~al.}(2009){Morganti}, {Peck}, {Oosterloo}, {van
  Moorsel}, {Capetti}, {Fanti}, {Parma}, \& {de Ruiter}}]{Morgantietal09}
{Morganti}, R., {Peck}, A.~B., {Oosterloo}, T.~A., {et~al.} 2009, \aap, 505,
  559, \dodoi{10.1051/0004-6361/200912605}

\bibitem[{{Morganti} {et~al.}(2006){Morganti}, {de Zeeuw}, {Oosterloo},
  {McDermid}, {Krajnovi{\'c}}, {Cappellari}, {Kenn}, {Weijmans}, \&
  {Sarzi}}]{Morgantietal06}
{Morganti}, R., {de Zeeuw}, P.~T., {Oosterloo}, T.~A., {et~al.} 2006, \mnras,
  371, 157, \dodoi{10.1111/j.1365-2966.2006.10681.x}

\bibitem[{{Morsony} {et~al.}(2010){Morsony}, {Heinz}, {Br{\"u}ggen}, \&
  {Ruszkowski}}]{Morsony2010}
{Morsony}, B.~J., {Heinz}, S., {Br{\"u}ggen}, M., \& {Ruszkowski}, M. 2010,
  \mnras, 407, 1277, \dodoi{10.1111/j.1365-2966.2010.17059.x}

\bibitem[{{Murgia} {et~al.}(2009){Murgia}, {Govoni}, {Markevitch}, {Feretti},
  {Giovannini}, {Taylor}, \& {Carretti}}]{Murgia2009}
{Murgia}, M., {Govoni}, F., {Markevitch}, M., {et~al.} 2009, \aap, 499, 679,
  \dodoi{10.1051/0004-6361/200911659}

\bibitem[{{Obreschkow} \& {Rawlings}(2009)}]{ObreschkowRawlings2009}
{Obreschkow}, D., \& {Rawlings}, S. 2009, \mnras, 394, 1857,
  \dodoi{10.1111/j.1365-2966.2009.14497.x}

\bibitem[{{Odekon} {et~al.}(2016){Odekon}, {Koopmann}, {Haynes}, {Finn},
  {McGowan}, {Micula}, {Reed}, {Giovanelli}, \& {Hallenbeck}}]{Odekonetal16}
{Odekon}, M.~C., {Koopmann}, R.~A., {Haynes}, M.~P., {et~al.} 2016, \apj, 824,
  110

\bibitem[{{Offringa} {et~al.}(2010){Offringa}, {de Bruyn}, {Biehl}, {Zaroubi},
  {Bernardi}, \& {Pandey}}]{Offringa2010}
{Offringa}, A.~R., {de Bruyn}, A.~G., {Biehl}, M., {et~al.} 2010, \mnras, 405,
  155, \dodoi{10.1111/j.1365-2966.2010.16471.x}

\bibitem[{{Offringa} {et~al.}(2014){Offringa}, {McKinley}, {Hurley-Walker},
  {Briggs}, {Wayth}, {Kaplan}, {Bell}, {Feng}, {Neben}, {Hughes}, {Rhee},
  {Murphy}, {Bhat}, {Bernardi}, {Bowman}, {Cappallo}, {Corey}, {Deshpand e},
  {Emrich}, {Ewall-Wice}, {Gaensler}, {Goeke}, {Greenhill}, {Hazelton},
  {Hindson}, {Johnston-Hollitt}, {Jacobs}, {Kasper}, {Kratzenberg}, {Lenc},
  {Lonsdale}, {Lynch}, {McWhirter}, {Mitchell}, {Morales}, {Morgan},
  {Kudryavtseva}, {Oberoi}, {Ord}, {Pindor}, {Procopio}, {Prabu}, {Riding},
  {Roshi}, {Shankar}, {Srivani}, {Subrahmanyan}, {Tingay}, {Waterson},
  {Webster}, {Whitney}, {Williams}, \& {Williams}}]{Offringa2014}
{Offringa}, A.~R., {McKinley}, B., {Hurley-Walker}, N., {et~al.} 2014, \mnras,
  444, 606, \dodoi{10.1093/mnras/stu1368}

\bibitem[{{Olivares} {et~al.}(2019){Olivares}, {Salome}, {Combes}, {Hamer},
  {Guillard}, {Lehnert}, {Polles}, {Beckmann}, {Dubois}, {Donahue}, {Edge},
  {Fabian}, {McNamara}, {Rose}, {Russell}, {Tremblay}, {Vantyghem}, {Canning},
  {Ferland}, {Godard}, {Peirani}, \& {Pineau des Forets}}]{Olivares2019}
{Olivares}, V., {Salome}, P., {Combes}, F., {et~al.} 2019, \aap, 631, A22,
  \dodoi{10.1051/0004-6361/201935350}

\bibitem[{{Olivares} {et~al.}(2022){Olivares}, {Salom{\'e}}, {Hamer}, {Combes},
  {Gaspari}, {Kolokythas}, {O'Sullivan}, {Beckmann}, {Babul}, {Polles},
  {Lehnert}, {Loubser}, {Donahue}, {Gendron-Marsolais}, {Lagos}, {Pineau des
  Forets}, {Godard}, {Rose}, {Tremblay}, {Ferland}, \&
  {Guillard}}]{Olivares2022}
{Olivares}, V., {Salom{\'e}}, P., {Hamer}, S.~L., {et~al.} 2022, \aap, 666,
  A94, \dodoi{10.1051/0004-6361/202142475}

\bibitem[{{O'Sullivan} {et~al.}(2014){O'Sullivan}, {David}, \&
  {Vrtilek}}]{O'Sullivan2014}
{O'Sullivan}, E., {David}, L.~P., \& {Vrtilek}, J.~M. 2014, \mnras, 437, 730,
  \dodoi{10.1093/mnras/stt1926}

\bibitem[{{O'Sullivan} {et~al.}(2018){O'Sullivan}, {Kolokythas}, {Kantharia},
  {Raychaudhury}, {David}, \& {Vrtilek}}]{OSullivanetal18}
{O'Sullivan}, E., {Kolokythas}, K., {Kantharia}, N.~G., {et~al.} 2018, \mnras,
  473, 5248, \dodoi{10.1093/mnras/stx2702}

\bibitem[{{O'Sullivan} {et~al.}(2007){O'Sullivan}, {Vrtilek}, {Harris}, \&
  {Ponman}}]{OSullivanetal07}
{O'Sullivan}, E., {Vrtilek}, J.~M., {Harris}, D.~E., \& {Ponman}, T.~J. 2007,
  \apj, 658, 299, \dodoi{10.1086/511778}

\bibitem[{{O'Sullivan} {et~al.}(2011){O'Sullivan}, {Worrall}, {Birkinshaw},
  {Trinchieri}, {Wolter}, {Zezas}, \& {Giacintucci}}]{OSullivanetal11}
{O'Sullivan}, E., {Worrall}, D.~M., {Birkinshaw}, M., {et~al.} 2011, \mnras,
  416, 2916, \dodoi{10.1111/j.1365-2966.2011.19239.x}

\bibitem[{{O'Sullivan} {et~al.}(2017){O'Sullivan}, {Ponman}, {Kolokythas},
  {Raychaudhury}, {Babul}, {Vrtilek}, {David}, {Giacintucci}, {Gitti}, \&
  {Haines}}]{OSullivanetal17}
{O'Sullivan}, E., {Ponman}, T.~J., {Kolokythas}, K., {et~al.} 2017, \mnras,
  472, 1482, \dodoi{10.1093/mnras/stx2078}

\bibitem[{{O'Sullivan} {et~al.}(2024){O'Sullivan}, {Rajpurohit},
  {Schellenberger}, {Vrtilek}, {David}, {Babul}, {Olivares}, {Ubertosi},
  {Kolokythas}, {Babyk}, \& {Loubser}}]{OSullivan2024}
{O'Sullivan}, E., {Rajpurohit}, K., {Schellenberger}, G., {et~al.} 2024, \apj,
  970, 65, \dodoi{10.3847/1538-4357/ad4ed6}

\bibitem[{{Pasini} {et~al.}(2021){Pasini}, {Finoguenov}, {Br{\"u}ggen},
  {Gaspari}, {de Gasperin}, \& {Gozaliasl}}]{Pasini2021}
{Pasini}, T., {Finoguenov}, A., {Br{\"u}ggen}, M., {et~al.} 2021, \mnras, 505,
  2628, \dodoi{10.1093/mnras/stab1451}

\bibitem[{{Pasini} {et~al.}(2022){Pasini}, {Br{\"u}ggen}, {Hoang},
  {Ghirardini}, {Bulbul}, {Klein}, {Liu}, {Shimwell}, {Hardcastle}, {Williams},
  {Botteon}, {Gastaldello}, {van Weeren}, {Merloni}, {de Gasperin}, {Bahar},
  {Pacaud}, \& {Ramos-Ceja}}]{Pasini2022}
{Pasini}, T., {Br{\"u}ggen}, M., {Hoang}, D.~N., {et~al.} 2022, \aap, 661, A13,
  \dodoi{10.1051/0004-6361/202141211}

\bibitem[{{Pl{\v{s}}ek} {et~al.}(2024){Pl{\v{s}}ek}, {Werner}, {Topinka}, \&
  {Simionescu}}]{Plsek2024}
{Pl{\v{s}}ek}, T., {Werner}, N., {Topinka}, M., \& {Simionescu}, A. 2024,
  \mnras, 527, 3315, \dodoi{10.1093/mnras/stad3371}

\bibitem[{{Pulido} {et~al.}(2018){Pulido}, {McNamara}, {Edge}, {Hogan},
  {Vantyghem}, {Russell}, {Nulsen}, {Babyk}, \& {Salom{\'e}}}]{Pulido2018}
{Pulido}, F.~A., {McNamara}, B.~R., {Edge}, A.~C., {et~al.} 2018, \apj, 853,
  177, \dodoi{10.3847/1538-4357/aaa54b}

\bibitem[{{Randall} {et~al.}(2015){Randall}, {Nulsen}, {Jones}, {Forman},
  {Bulbul}, {Clarke}, {Kraft}, {Blanton}, {David}, {Werner}, {Sun}, {Donahue},
  {Giacintucci}, \& {Simionescu}}]{Randalletal15}
{Randall}, S.~W., {Nulsen}, P.~E.~J., {Jones}, C., {et~al.} 2015, \apj, 805,
  112, \dodoi{10.1088/0004-637X/805/2/112}

\bibitem[{{Robitaille} \& {Bressert}(2012)}]{aplpy}
{Robitaille}, T., \& {Bressert}, E. 2012, {APLpy: Astronomical Plotting Library
  in Python}.
\newblock \doeprint{1208.017}

\bibitem[{{Rose} {et~al.}(2019){Rose}, {Edge}, {Combes}, {Gaspari}, {Hamer},
  {Nesvadba}, {Peck}, {Sarazin}, {Tremblay}, {Baum}, {Bremer}, {McNamara},
  {O'Dea}, {Oonk}, {Russell}, {Salom{\'e}}, {Donahue}, {Fabian}, {Ferland},
  {Mittal}, \& {Vantyghem}}]{Rose2019}
{Rose}, T., {Edge}, A.~C., {Combes}, F., {et~al.} 2019, \mnras, 489, 349,
  \dodoi{10.1093/mnras/stz2138}

\bibitem[{{Rose} {et~al.}(2023){Rose}, {McNamara}, {Combes}, {Edge}, {Fabian},
  {Gaspari}, {Russell}, {Salom{\'e}}, {Tremblay}, \& {Ferland}}]{Rose2023}
{Rose}, T., {McNamara}, B.~R., {Combes}, F., {et~al.} 2023, \mnras, 518, 878,
  \dodoi{10.1093/mnras/stac3194}

\bibitem[{{Rose} {et~al.}(2024){Rose}, {McNamara}, {Combes}, {Edge},
  {McDonald}, {O'Sullivan}, {Russell}, {Fabian}, {Ferland}, {Salom{\'e}}, \&
  {Tremblay}}]{Rose2024}
---. 2024, \mnras, 533, 771, \dodoi{10.1093/mnras/stae1831}

\bibitem[{{Ruffa} {et~al.}(2019{\natexlab{a}}){Ruffa}, {Prandoni}, {Laing},
  {Paladino}, {Parma}, {de Ruiter}, {Mignano}, {Davis}, {Bureau}, \&
  {Warren}}]{Ruffa2019}
{Ruffa}, I., {Prandoni}, I., {Laing}, R.~A., {et~al.} 2019{\natexlab{a}},
  \mnras, 484, 4239, \dodoi{10.1093/mnras/stz255}

\bibitem[{{Ruffa} {et~al.}(2019{\natexlab{b}}){Ruffa}, {Davis}, {Prandoni},
  {Laing}, {Paladino}, {Parma}, {de Ruiter}, {Casasola}, {Bureau}, \&
  {Warren}}]{Ruffa2019a}
{Ruffa}, I., {Davis}, T.~A., {Prandoni}, I., {et~al.} 2019{\natexlab{b}},
  \mnras, 489, 3739, \dodoi{10.1093/mnras/stz2368}

\bibitem[{{Russell} {et~al.}(2019){Russell}, {McNamara}, {Fabian}, {Nulsen},
  {Combes}, {Edge}, {Madar}, {Olivares}, {Salom{\'e}}, \&
  {Vantyghem}}]{Russelletal19}
{Russell}, H.~R., {McNamara}, B.~R., {Fabian}, A.~C., {et~al.} 2019, \mnras,
  490, 3025, \dodoi{10.1093/mnras/stz2719}

\bibitem[{{Salom{\'e}} \& {Combes}(2003)}]{SalomeCombes2003}
{Salom{\'e}}, P., \& {Combes}, F. 2003, \aap, 412, 657,
  \dodoi{10.1051/0004-6361:20031438}

\bibitem[{{Sanders} {et~al.}(2009){Sanders}, {Fabian}, \&
  {Taylor}}]{Sanders2009}
{Sanders}, J.~S., {Fabian}, A.~C., \& {Taylor}, G.~B. 2009, \mnras, 396, 1449,
  \dodoi{10.1111/j.1365-2966.2009.14892.x}

\bibitem[{{Saraf} {et~al.}(2023){Saraf}, {Wong}, {Cortese}, \&
  {Koribalski}}]{Sarafetal23}
{Saraf}, M., {Wong}, O.~I., {Cortese}, L., \& {Koribalski}, B.~S. 2023, \mnras,
  519, 4128, \dodoi{10.1093/mnras/stac3695}

\bibitem[{{Scaife} \& {Heald}(2012)}]{Scaife2012}
{Scaife}, A. M.~M., \& {Heald}, G.~H. 2012, \mnras, 423, L30,
  \dodoi{10.1111/j.1745-3933.2012.01251.x}

\bibitem[{{Schellenberger} {et~al.}(2021){Schellenberger}, {David}, {Vrtilek},
  {O'Sullivan}, {Giacintucci}, {Forman}, {Jones}, \&
  {Venturi}}]{Schellenberger2021}
{Schellenberger}, G., {David}, L.~P., {Vrtilek}, J., {et~al.} 2021, \apj, 906,
  16, \dodoi{10.3847/1538-4357/abc488}

\bibitem[{{Schellenberger} {et~al.}(2020){Schellenberger}, {David}, {Vrtilek},
  {O'Sullivan}, {Lim}, {Forman}, {Sun}, {Combes}, {Salome}, {Jones},
  {Giacintucci}, {Edge}, {Gastaldello}, {Temi}, {Brighenti}, \&
  {Bardelli}}]{Schellenberger2020}
---. 2020, \apj, 894, 72, \dodoi{10.3847/1538-4357/ab879c}

\bibitem[{{Schellenberger} {et~al.}(2024){Schellenberger}, {O'Sullivan},
  {David}, {Vrtilek}, {Romero}, {Petitpas}, {Forman}, {Giacintucci}, {Gurwell},
  {Jones}, {Rajpurohit}, {Ubertosi}, \& {Venturi}}]{Schellenberger2024}
{Schellenberger}, G., {O'Sullivan}, E., {David}, L.~P., {et~al.} 2024, \apj,
  976, 246, \dodoi{10.3847/1538-4357/ad89bc}

\bibitem[{{Serra} {et~al.}(2012){Serra}, {Oosterloo}, {Morganti}, {Alatalo},
  {Blitz}, {Bois}, {Bournaud}, {Bureau}, {Cappellari}, {Crocker}, {Davies},
  {Davis}, {de Zeeuw}, {Duc}, {Emsellem}, {Khochfar}, {Krajnovi{\'c}},
  {Kuntschner}, {Lablanche}, {McDermid}, {Naab}, {Sarzi}, {Scott}, {Trager},
  {Weijmans}, \& {Young}}]{Serra2012}
{Serra}, P., {Oosterloo}, T., {Morganti}, R., {et~al.} 2012, \mnras, 422, 1835,
  \dodoi{10.1111/j.1365-2966.2012.20219.x}

\bibitem[{{Serra} {et~al.}(2015){Serra}, {Westmeier}, {Giese}, {Jurek},
  {Fl{\"o}er}, {Popping}, {Winkel}, {van der Hulst}, {Meyer}, {Koribalski},
  {Staveley-Smith}, \& {Courtois}}]{Serra2015}
{Serra}, P., {Westmeier}, T., {Giese}, N., {et~al.} 2015, \mnras, 448, 1922,
  \dodoi{10.1093/mnras/stv079}

\bibitem[{{Temi} {et~al.}(2018){Temi}, {Amblard}, {Gitti}, {Brighenti},
  {Gaspari}, {Mathews}, \& {David}}]{Temi2018}
{Temi}, P., {Amblard}, A., {Gitti}, M., {et~al.} 2018, \apj, 858, 17,
  \dodoi{10.3847/1538-4357/aab9b0}

\bibitem[{{Tonry} {et~al.}(2001){Tonry}, {Dressler}, {Blakeslee}, {Ajhar},
  {Fletcher}, {Luppino}, {Metzger}, \& {Moore}}]{Tonry2001}
{Tonry}, J.~L., {Dressler}, A., {Blakeslee}, J.~P., {et~al.} 2001, \apj, 546,
  681, \dodoi{10.1086/318301}

\bibitem[{{Ubertosi} {et~al.}(2023){Ubertosi}, {Gitti}, {Brighenti},
  {McDonald}, {Nulsen}, {Donahue}, {Brunetti}, {Randall}, {Gaspari}, {Ettori},
  {Calzadilla}, {Ignesti}, {Feretti}, \& {Blanton}}]{Ubertosietal23}
{Ubertosi}, F., {Gitti}, M., {Brighenti}, F., {et~al.} 2023, \apj, 944, 216,
  \dodoi{10.3847/1538-4357/acacf9}

\bibitem[{{van Weeren} {et~al.}(2024){van Weeren}, {Timmerman}, {Vaidya},
  {Gendron-Marsolais}, {Botteon}, {Roberts}, {Hlavacek-Larrondo}, {Bonafede},
  {Br{\"u}ggen}, {Brunetti}, {Cassano}, {Cuciti}, {Edge}, {Gastaldello},
  {Groeneveld}, \& {Shimwell}}]{vanWeeren2024}
{van Weeren}, R.~J., {Timmerman}, R., {Vaidya}, V., {et~al.} 2024, arXiv
  e-prints, arXiv:2410.02863, \dodoi{10.48550/arXiv.2410.02863}

\bibitem[{{Vantyghem} {et~al.}(2017){Vantyghem}, {McNamara}, {Edge}, {Combes},
  {Russell}, {Fabian}, {Hogan}, {McDonald}, {Nulsen}, \&
  {Salom{\'e}}}]{Vantyghem2017}
{Vantyghem}, A.~N., {McNamara}, B.~R., {Edge}, A.~C., {et~al.} 2017, \apj, 848,
  101, \dodoi{10.3847/1538-4357/aa8fd0}

\bibitem[{{Welch} {et~al.}(2010){Welch}, {Sage}, \& {Young}}]{Welch2010}
{Welch}, G.~A., {Sage}, L.~J., \& {Young}, L.~M. 2010, \apj, 725, 100,
  \dodoi{10.1088/0004-637X/725/1/100}

\bibitem[{{Werner} {et~al.}(2014){Werner}, {Oonk}, {Sun}, {Nulsen}, {Allen},
  {Canning}, {Simionescu}, {Hoffer}, {Connor}, {Donahue}, {Edge}, {Fabian},
  {von der Linden}, {Reynolds}, \& {Ruszkowski}}]{Werner2014}
{Werner}, N., {Oonk}, J.~B.~R., {Sun}, M., {et~al.} 2014, \mnras, 439, 2291,
  \dodoi{10.1093/mnras/stu006}

\bibitem[{{Wolfire} {et~al.}(1995){Wolfire}, {Hollenbach}, {McKee}, {Tielens},
  \& {Bakes}}]{Wolfire1995}
{Wolfire}, M.~G., {Hollenbach}, D., {McKee}, C.~F., {Tielens}, A.~G.~G.~M., \&
  {Bakes}, E.~L.~O. 1995, \apj, 443, 152, \dodoi{10.1086/175510}

\bibitem[{{Young} {et~al.}(2011){Young}, {Bureau}, {Davis}, {Combes},
  {McDermid}, {Alatalo}, {Blitz}, {Bois}, {Bournaud}, {Cappellari}, {Davies},
  {de Zeeuw}, {Emsellem}, {Khochfar}, {Krajnovi{\'c}}, {Kuntschner},
  {Lablanche}, {Morganti}, {Naab}, {Oosterloo}, {Sarzi}, {Scott}, {Serra}, \&
  {Weijmans}}]{Young2011}
{Young}, L.~M., {Bureau}, M., {Davis}, T.~A., {et~al.} 2011, \mnras, 414, 940,
  \dodoi{10.1111/j.1365-2966.2011.18561.x}

\end{thebibliography}

\end{document}